\documentclass[final,1p,twocolumn]{elsarticle}
  \usepackage{graphicx}
  \usepackage{amsmath,amssymb}
  \usepackage{subfigure}
  \usepackage{color}

  \newcommand{\dif}{\mathrm{d}}
  \newcommand{\J}{\mathcal{J}}

  \newcommand{\x}{{\bf x}}

  \newtheorem{problem}{Problem}

 \usepackage{style}

  \journal{-}

\begin{document}

  \begin{frontmatter}
  
 \title{Data Assimilation for Navier-Stokes using the Least-Squares
    Finite-Element Method}

  \author{Alexander Schwarz}
  \address{Institut f\"ur Mechanik, University of Duisburg-Essen, 
    Universit\"atsstra{\ss}e 15, 45117 Essen, Germany.}
  \author{Richard P. Dwight}
  \address{Aerodynamics Group, Faculty of Aerospace,
            TU Delft, P.O.\ Box 5058, 2600GB Delft, The Netherlands.}
  
  \begin{abstract}
    We investigate theoretically and numerically the use of the Least-Squares
    Finite-element method (LSFEM) to approach data-assimilation problems for the
    steady-state, incompressible Navier-Stokes equations.  Our LSFEM
    discretization is based on a stress-velocity-pressure (S-V-P) first-order
    formulation, using discrete counterparts of the Sobolev spaces $H(\div)
    \times H^1 \times L^2$ for the variables respectively.  In general S-V-P
    formulations are promising when the stresses are of special interest,
    e.g.~for non-Newtonian, multiphase or turbulent flows.
    Resolution of the system is via minimization of a least-squares functional
    representing the magnitude of the residual of the equations.  A simple and
    immediate approach to extend this solver to data-assimilation is to add a
    data-discrepancy term to the functional.  Whereas most data-assimilation
    techniques require a large number of evaluations of the forward-simulation
    and are therefore very expensive, the approach proposed in this work 
    uniquely has the same cost
    as a single forward run.  However, the question arises: what is the
    statistical model implied by this choice?  We answer this within the
    Bayesian framework, establishing the latent background covariance model and
    the likelihood.  Further we demonstrate that -- in the linear case -- the
    method is equivalent to application of the Kalman filter, and derive the
    posterior covariance.  We practically demonstrate the capabilities of our
    method on a backward-facing step case.  Our LSFEM formulation (without data)
    is shown to have good approximation quality, even on relatively coarse
    meshes - in particular with respect to mass-conservation and reattachment
    location.  Adding limited velocity measurements from experiment, we show
    that the method is able to correct for discretization error on very coarse
    meshes, as well as correct for the influence of unknown and uncertain
    boundary-conditions.
  \end{abstract}
  

\end{frontmatter}

\clearpage

%%%%%%%%%%%%%%%%%%%%%%%%%%%%%%%%%%%%%%%%%%%%%%%%%%%%%%%%%%%
\section{Introduction}
\label{s_intro}
Increasingly there is a demand for numerical methods for fluids capable of
assimilating experimental data from wind-tunnel measurements into CFD
simulation.  This can be done with several goals:
identifying unmeasured variables such as pressure from particle-image
velocimetry (PIV) velocity field measurements~\cite{Rag2012}; increasing the
time-resolution of measurement data~\cite{ScaMoo2012}; filling spatial gaps in
PIV data~\cite{SchDwiSca2012}; or filtering noisy data~\cite{ConAni2012}.  The methods used may be seen as physics-based
filtering/regression techniques.  In each case the ability of a simulation to smoothly
approximate the flow-field in the full domain in all variables, or identify an
unmeasured quantity of interest, increases the value of the measurement.  An
increase in the amount of data provided by modern experimental techniques drives
a trend toward increasing demand for effective post-processing tools, with an
increasing amount of physical
fidelity~\cite{RagVouSca2011,ScaMoo2012,SchDwiSca2012}.

In this paper we investigate an approach to this problem for assimilating
experimental velocity measurements into Navier-Stokes flow based on a
Least-Squares Finite-Element Method (LSFEM)~\cite{Jia1998,BocGun2009}.  Briefly:
given a system of PDEs with boundary conditions, LSFEM methods define a scalar
functional whose minimum is the solution of the system.  The functional is
defined as a sum of e.g.\ $L^2$-norms of the residuals of the original PDE and
optionally BCs.  These terms are then discretized with finite elements spaces.
By construction at least one minimum of the functional exists, even when the
original PDE plus BCs is ill-posed in the sense of Hadamard.  Since the
discretization is already constructed in a least-squares sense, to add
information from experimental data, we only have to add a term quadratically
penalizing the difference between predictions and observations.  The weighting
of this term determines the extent to which the solution will match the
observations and is chosen in inverse proportion to the measurement accuracy.

The method (which we refer to as LSFEM+) was first proposed for Navier-Stokes by
Heyes et al.\ \cite{HeyManMcc2010} and later investigated by the present
authors~\cite{Dwi2010}.  It has desirable theoretical behaviour in the limits:
in the absence of data we retrieve the standard LSFEM flow solution, and for
plentiful, perfectly accurate data we interpolate the data exactly.  Furthermore
LSFEM+ is unusual in being capable of data assimilation at a computational
cost equivalent to a single flow simulation, see Section~\ref{ss:dacontext}.

The accuracy of the method will depend on the quality of the underlying LSFEM
discretization in regions in which no data is available.  Our aim in this paper
is to provide a mixed LSFEM with improved accuracy (also with respect to mass
conservation), but without additional variables or degrees of freedom.  Starting from the
incompressible Navier-Stokes equations we derive a div-grad first-order system
resulting in a three-field approach with stresses, velocities, and pressure as
unknowns.  In our formulation the stresses are approximated by using shape
functions related to the edges. These quadratic vector-valued functions are used
for the interpolation of the rows of the stress tensor and belong to a
Raviart-Thomas space, which guarantees a conforming discretization of the
Sobolev space $H(div)$.  Furthermore, standard cubic and linear polynomials
associated to the vertices of the triangle are used for the continuous
approximation of the velocities and the pressure respectively.  This approach is
similar that of Cai et al.~\cite{CaiLeeWan:2004:lsm} who developed theoretical
properties of this discretization for Stokes flow.

In summary, the contributions of this paper are: firstly to extend the
stress-velocity-pressure formulation to Navier-Stokes, and perform a numerical
investigation of a real experiment.  Secondly to show the effectiveness and
efficiency of data assimilation with this LSFEM approach, including solving
inverse problems with uncertain and unknown boundary conditions, at a
computational cost equivalent to a single solve of the forward problem.  Thirdly
to motivative the choice of data weighting based on measurement error in a
Bayesian statistical framework.

The paper is outlined as follows: in Section~\ref{ss:dacontext} we give a short
overview of alternative data assimilation approaches, and in
Section~\ref{ss:history} a brief overview of LSFEM methods for fluids.  The
stress-velocity-pressure formulation is derived in Section~\ref{s:lsfem}.  The
Bayesian derivation and analysis of the added-data method is given in
Section~\ref{s:bayes}.  The method is numerically evaluated for a laminar
backward-facing step in Section~\ref{s_fuckedifiknow}, and conclusions are given
in Section~\ref{s_conclusions}.

%%%%%%%%%%%%%%%%%%%%%%%%%%%%%%%%%%%%%%%%%%%%%%%%%%%%%%%%%%%%%%%%%%%%%%%%%%%%%%%%
\subsection{Data Assimilation Context}
\label{ss:dacontext}
As mentioned LSFEM+ has a cost of a single forward flow simulation.  In contrast
the computational cost of some standard stochastic data assimilation methods is
much higher: Ensemble Kalman filtering costs typically 100 flow
solves~\cite{Eve2003,EveLee2000}; 4D-Var requires about 50-100 flow and adjoint
solves for nonlinear problems; physics based Reduced Order Model fitting (ROMs)
requires training the ROM on the parameter space of interest, for rarely less
than 100 solves~\cite{LegAlo2004,ZimGoe2012}.

There are two other notable approaches with a comparable computational cost to
our proposed method.  The first is Optimal Interpolation, see
e.g.~\cite{Eve2003}, a minimum variance estimator.  It is similar in form to the
Kalman filter, in that it consists of a forecast step followed by an analysis
step, except that the forecast step does not propagate the uncertainty through
the model as in the Kalman filter, instead only the mean is propagated and fresh
randomness is generated on the result according to a user-specified covariance
structure.  Thus the magnitude and covariance structure of the uncertainty due
to the model is not preserved, and thus the method relies on a good guess of
these.  The second ``cheap'' approach is Measurement-Integrated
Simulation~\cite{HayImaFun2010,ImaHay2010}, in which a non-physical forcing term
is added to the Navier-Stokes equations.  The forcing is designed to push the
solution in the direction of experimental observations, and disappears in the
case of exact experiment-simulation agreement.  The strength of the forcing is
an influential free parameter, having a similar effect to the Kalman gain in the
Kalman filter, but currently the statistical interpretation of this approach is
not clear.
%
% The current application of LSFEM to data assimilation has some
% relation to control problems
% \item LSFEM for control problems \cite{BedFix1995,Boc1997,GunLee2000}

%%%%%%%%%%%%%%%%%%%%%%%%%%%%%%%%%%%%%%%%%%%%%%%%%%%%%%%%%%%%%%%%%%%%%%%%%%%%%%%%
\subsection{Least-Squares Finite-Elements for Incompressible Navier-Stokes}
\label{ss:history}
In recent years the least-squares method was successfully applied to many
problems in fluid dynamics and solid mechanics.  The development and improvement
of mixed least-squares finite element formulations have been the objective of
many studies.  Exhaustive overviews with respect to theoretical foundations and
application-oriented issues can be found for example in the
monographs~\cite{Jia:1998:tls} and \cite{BocGun:2009:lsf} and in the
review~\cite{KayMat:2005:lsf}.

Least-squares variational principles have increasingly gained attention, which
is due to some theoretical and computational advantages compared to the Galerkin
method. But nevertheless, the Galerkin method is today the most employed
variational approach for all kind of boundary value problems in computational
mechanics.  The least-squares method in contrast only play a minor role.  From
the point of view of an engineer the reasons are evident: since the 1970s
attention has focussed on low-order elements which are easy to implement,
robust, and lead to sparse, well-conditioned stiffness matrices.  Unfortunately,
the approximation quality of low-order least-squares finite elements is moderate and
consequently only little attention was paid to the method.  An illustrative
example for the poor solution of lower-order interpolation was given in
\cite{pon:2003:lsv} for the driven cavity problem.  While it is possible to
obtain qualitatively correct results, the method consistently needs
more degrees of freedom compared to alternative methods.  Despite this
drawback, the least-squares method has also some important advantages
\cite{Jia:1998:tls,BocGun:2009:lsf}
\begin{itemize}
\item The method provides an {\it a posteriori} error estimate
      without additional costs, which can be used for adaptive mesh refinement.
\item The flexibility to design functionals directly approximating
  field variables of interest, e.g. stresses, vorticity, velocities.
\item The method is not restricted by the LBB-condition regarding the
  choice of the polynomial degree of the finite element spaces.
\item The resulting positive definite system matrices can be solved by
  using robust and fast iterative methods even for problems with
  non-self-adjoint operators.\\[-0.0cm]
\end{itemize}
These features yield the basis for theoretical analysis and numerical studies on
e.g.\ ellipticity, error estimation, residual weighting and convergence and are
investigated in the field of fluid dynamics by many authors, see e.g. 
\cite{BocGun:2009:lsf} and the references therein.
The utilized first-order systems in general consist of a composition of different
quantities and could be divided into three approaches:
\begin{itemize}
\item The vorticity, velocity, pressure (VVP) formulation, see e.g.\  \cite{Jia:1998:tls}, \cite{JiaCha:1990:lsf} and \cite{Boc:1997:aol}.
\item The stresses, velocity, pressure (SVP) formulation, see e.g.\ \cite{BelSur:1994:pls}, \cite{BocGun:1995:lsm} and \cite{DinTsa:2003:ofo}.
\item The velocity gradient, velocity, pressure (VGVP) formulation, see e.g.\ \cite{CaiManMcC:1997:fos}, \cite{BocCaiManMcC:1998:aov} and \cite{BocManMcC:1999:aov}.
\end{itemize}

The utilized first-order systems in general consist of a composition of the flow
velocity, the gradient of the flow velocity, the pressure, the stresses, and the
vorticity as dependent variables.

The main point of criticism of LSFEM in fluid-dynamics is the lack of numerical
conservation, in particular mass conservation.  Various strategies, e.g.\ mesh
refinement, residual weighting, use of Lagrange multipliers, and improved
velocity-pressure coupling have been used to mitigate this disadvantage somewhat
\cite{BocManMcC:1999:aov,DeaGun:1998:irt,BolTha:2005:omc,ChaNel:1997:lsf,Pon:2006:als}.
In recent works, spectral finite elements with higher-order polynomial finite
element spaces ($p \geq 4 $) have demonstrated a satisfying approximation of the
governing fields, with improved mass conservation 
\cite{ponred:2003:sls,ponred:2004:stc,ProGer:2006:mam}.  Extensive work was done
by the group from the University of Boulder
\cite{HeyLeeManMcC:2006:omc,HeyLeeManMcC:2007:aal,HeyLeeManMcCRug:2009:emc} on
the choice of first-order system, aiming to achieve stronger coupling between
velocity and pressure, realised by introducing new variables depending on both
fields, or by special treatment of the boundary conditions.  The benefits
achieved by all these techniques in combination allows acceptable performance of
LSFEM for incompressible fluids.

%%%%%%%%%%%%%%%%%%%%%%%%%%%%%%%%%%%%%%%%%%%%%%%%%%%%%%%%%%%%%%%%%%%%%%%%%%%%%%%%
\section{LSFEM Discretization of Navier-Stokes}
\label{s:lsfem}
In the following we describe the least-squares mixed finite element method for
first-order linear and nonlinear systems, see e.g. \cite{Jia1998,BocGun2009}.  A
stress-velocity-pressure formulation for the incompressible Navier-Stokes equations is constructed, and
finally we introduce a data term into the LSFEM functional.  We work on the open
bounded domain $\Omega\subset\mathbb{R}^d$ with piecewise-smooth boundary
$\Gamma$.

\subsection{Linear PDEs}
Consider a (first-order) system of linear partial differential equations
\begin{equation}
Lw=f\:\:\text{on $\Omega$}, \qquad Bw=g\:\:\text{on $\Gamma$},
\label{eq:lingov}
\end{equation}
where $f\in L^2(\Omega)$, $g\in L^2(\Gamma)$ and $L$ and $B$ denote linear
differential operators on $\Omega$ and $\Gamma$ respectively.  Let
$(\cdot,\cdot)_\Omega$ and $(\cdot,\cdot)_\Gamma$ represent the $L^2(\Omega)$
and $L^2(\Gamma)$ scalar products.  Also let $V$ be a suitable Sobolev space,
then \eqref{eq:lingov} can be reformulated as~\cite{Jia1998}
\[
 \min_{w\in V}\underbrace{\left[\|Lw - f\|^2_\Omega + \|Bw-g\|^2_\Gamma\right]}_{J(w;f,g)},
\]
where $\|\cdot\|^2_\Omega$ and $\|\cdot\|^2_\Gamma$ are the norms induced by the above
scalar products, and $J$ is the {\it least-squares functional}.

The least-squares variational formulation is obtained by requiring that the
first variation of $J$ vanishes at $w$ for all admissible perturbuations $\delta w\in V$:
\[
\lim_{\epsilon\rightarrow 0} \frac{\dif}{\dif\epsilon} J(w +\epsilon \delta w) = 0,
\]
from which follows: Find $w\in V$ such that
\eb
\underbrace{(Lw, L\delta w)_\Omega + (Bw, B\delta w)_\Gamma}_{Q(w,\delta w)} = 
  \underbrace{(f,L\delta w)_\Omega + (g,B\delta w)_\Gamma}_{l(\delta w)},\quad 
\forall \delta w\in V,
\label{eq:varpart}
\ee 
where $Q(\cdot,\cdot)$ is a symmetric bilinear form and $l(\cdot)$ a linear
form.  This equation can be discretized by replacing $V$ with a discrete
function space $V_h\subset V$.  The resulting discretization can be shown to
have a unique stable solution provided $L$ is bounded below~\cite{BocGun2009} --
in particular no equivalent of the LBB condition from mixed Galerkin
formulations is required.

\subsection{Nonlinear PDEs}
Consider the nonlinear system of PDEs
\begin{equation}
Nw=0\quad\text{on $\Omega$}, \qquad Bw=0\quad\text{on $\Gamma$},
\label{eq:nonlingov}
\end{equation}
where $N$ is a nonlinear differential operator on $\Omega$, and $B$ is a
nonlinear boundary operator. Assume we can find a suitable function space $V$
for solutions of this system.  There are two options for formulating nonlinear
LSFEM, we can either (a) linearize $N$, and build the LSFEM functional on the
linearized governing equation, which then acts as a Newton iteration for the nonlinear
system, or (b) build the functional directly on the nonlinear operator, and use
Newton's method to find a root of the 1st-variation~\cite{Kay:2006:lsm}.
The latter results in an extra term in the 2nd-variation; and this is the
approach we follow here.

The LSFEM formualtion is then
\[
\min_{w\in W} \left[\| Nw \|_\Omega^2 + \| Bw \|_\Gamma^2 \right] = 
\min_{w\in W} \underbrace{\left[(Nw,Nw)_\Omega + (Bw,Bw)_\Gamma \right]}_{\mathcal{J}(w)}.
\]
Taking 1st-variations gives
\[
\delta \mathcal{J}(w; \delta w) = (Nw, N'\delta w)_\Omega + (Bw, B'\delta w)_\Gamma,
\]
assuming that the Fr\'echet derivatives $N'[w]$, $B'[w]$ of $N$ and $B$ about
$w$ exist.  Second-variations results in
\begin{align*}
\Delta \delta \mathcal{J}(w; \delta w, \Delta w) &= (N'\Delta w, N'\delta w)_\Omega + 
                               (Nw, N''\Delta \delta w)_\Omega\\ &+ 
                               (B'\Delta w, B'\delta w)_\Gamma +
                               (Bw, B''\Delta \delta w)_\Gamma,
\end{align*}
where the final term is zero in the common case of linear $B$.  Note that the
term $(Nw, N''\Delta \delta w)_\Omega$ is not present in the nonlinear
formulation (a) mentioned above.  Given these variations the Newton iteration
for the update $\Delta w$ is:
\[
\Delta \delta \mathcal{J}(w; \delta w, \Delta w) + \delta \mathcal{J}(w; \delta w) = 0.
\]

\subsection{Navier-Stokes in $\Bsigma-\bv-p$ form}
In order to describe a boundary-value problem for Newtonian fluid flow, we
consider the steady incompressible Navier-Stokes equations consisting of
balances of momentum and mass 
\eb 
\begin{array}{rcl}
- \rho \nabla \bv \, \bv + 2 \rho \nu  \, \mbox{div}  \, 
\nabla^s \bv - \nabla p &=&  \bzero \\ 
\mbox{div} \, \bv &=&  0 \; .
\end{array}
\label{bal}
\ee
Here, $\bv$ denotes the velocities, $p$ the pressure, $\rho$ the density
and $\nu$ the kinematic viscosity of a medium flowing through the domain. 
Furthermore,
\eb
\nabla^s \bv = \dfrac {1}{2} \left( \nabla \bv + \nabla \bv^T \right)
\ee
denotes the symmetric velocity gradient. System (\ref{bal}) is
of second order and can be transformed into a first-order div-grad system
by means of an additional variable, namely the Cauchy stresses
\eb
\Bsigma = 2 \rho \nu \nabla^s \bv - p \bone \; .
\ee
We achieve
\eb
\begin{array}{rcl}
\mbox{div} \, \Bsigma - \rho \nabla \bv \, \bv &=&  \bzero \\ 
\Bsigma - 2 \rho \nu \nabla^s \bv + p \bone &=&  \bzero \\ 
\mbox{div} \, \bv &=&  0 \; ,
\end{array}
\label{svp}
\ee
which is the basis for the nonlinear differential operator $N$, see
e.g.~\cite{CaiLeeWan:2004:lsm}.  In order to complete the boundary value problem
we split the boundary of the domain $\Gamma$ into two subsets, namely the
Dirichlet boundary $\Gamma_D$ and Neumann boundary $\Gamma_N$ with the
definitions
\eb
\Gamma_D \cup \Gamma_N = \Gamma \qquad \mbox{and} \qquad
\Gamma_D \cap \Gamma_N = \emptyset \, ,
\ee
on which the boundary conditions are
\eb
\bv = \bg  \quad \mbox{on} \; \Gamma_D \; , \qquad
p = q  \qquad \mbox{and} \qquad
\Bsigma \bn = \bh  \quad \mbox{on} \; \Gamma_N \; .
\ee
The resulting quadratic least-squares functional for the set of
equations (\ref{svp}) is given in terms of the $L_2$-norm as
\eb
\begin{array}{rcl}
\J (\Bsigma,\bv,p) = 
\dfrac{1}{2} ( \| \dfrac{1}{\sqrt{\rho}}(\mbox{div} \, \Bsigma - \rho \nabla \bv \, \bv)  \|^2
+ \| \dfrac{1}{\sqrt{\rho \nu}} (\Bsigma - 2 \rho \nu \nabla^s \bv + p \bone) \|^2 
+ \| \mbox{div} \, \bv \|^2) \; ,
\end{array}
\label{funcall}
\ee
where we have used some additional physically motivated 
weights on the 1st and 2nd terms, see e.g. \cite{SchNicSerNisOuaSchTur:2016:acs}.

The approximation spaces for 
$\J(\Bsigma, \bv, p) :  \bX \times \bV \times \bQ \rightarrow \IR$
are given by
\ebn
\displaystyle
\bX = \left\{ \Bsigma \in H(\mbox{div},\Omega)^2 \right\}
\; , \qquad \bV = \left\{ \bv \in H^1(\Omega)^2 \right\} 
\qquad \mbox{and} \qquad
\bQ = \left\{ p \in L_2(\Omega)^2 \right\}.
\een
The minimization of $\J$ presupposes the first variation is zero
\eb
\delta_{\sigma, v, p} \J (\Bsigma, \bv, p, \delta \Bsigma, \delta \bv, \delta p) = 0
\ee
or explicitly:
\eb
\renewcommand{\arraystretch}{2.0}
\begin{array}{rcl}
\delta_{v} \J  &=&   \displaystyle  
   \int_{\Omega} \mbox{div} \, \delta \bv \cdot \mbox{div} \, \bv \, dV
 - \displaystyle  \dfrac{1}{\rho \nu} \, \int_{\Omega} 2 \rho \nu \nabla^s \, \delta \bv 
 \cdot \left( \Bsigma - 2 \rho \nu \nabla^s \bv + p {\bf 1} \right) \, dV\\
&&-  \displaystyle \dfrac{1}{\rho} \, \int_{\Omega} 
 \left(\rho \nabla \delta \bv\, \bv  + \rho \nabla \bv \, \delta \bv  \right)  
 \cdot \left( \mbox{div} \, \Bsigma  - \rho \nabla \bv \, \bv \right) \, dV = 0\\[0.3cm]
\delta_{\sigma} \J &=&   \displaystyle
   \dfrac{1}{\rho} \, \int_{\Omega} \mbox{div} \, \delta \Bsigma 
  \cdot \left( \mbox{div} \, \Bsigma - \rho \nabla \bv \, \bv \right) \, dV 
  + \displaystyle \dfrac{1}{\rho \nu} \, \int_{\Omega} \delta \Bsigma 
  \cdot \left( \Bsigma - 2 \rho \nu \nabla^s \bv + p {\bf 1} \right) \, dV = 0 \\[0.3cm]
\delta_{p} \J &=&   \displaystyle
   \dfrac{1}{\rho \nu} \, \int_{\Omega} \delta p {\bf 1} 
  \cdot \left( \Bsigma - 2 \rho \nu \nabla^s \bv + p {\bf 1} \right) \, dV = 0 \; .
\end{array}
\label{first}
\ee
In order to seek the minimizer with Newton's method, we linearize $\delta \J$
with respect to stresses, velocities and pressure.  Consequently, the second
variation of $\J$ denoted as $\Delta\delta \J$, which is the basis for the
Newton tangent, is explicitly
\eb
\renewcommand{\arraystretch}{2.0}
\begin{array}{rcl}
\Delta \delta_{vv} \J  &=&  \displaystyle  
   \int_{\Omega} \mbox{div} \, \delta \bv \cdot \mbox{div} \, \Delta \bv \, dV
 + \displaystyle \dfrac{1}{\rho \nu} \, \int_{\Omega}  2 \rho \nu \nabla^s \, \delta \bv 
                 \cdot 2 \rho \nu \nabla^s \, \delta \bv  \, dV \\
&&+ \displaystyle \dfrac{1}{\rho} \, \int_{\Omega}  
\left(\rho \nabla \delta \bv \, \bv + \rho \nabla \bv \, \delta \bv  \right)  
      \cdot 
\left(\rho \nabla \Delta \bv \, \bv + \rho \nabla \bv \, \Delta \bv \right) \, dV \\
&&-  \displaystyle \dfrac{1}{\rho} \, \int_{\Omega} 
  \left( \rho \nabla \delta \bv\, \Delta \bv + \rho \nabla \Delta \bv \, \delta \bv  \right)  
      \cdot \left( \mbox{div} \, \Bsigma - \rho \nabla \bv \, \bv \right) \, dV \\
\Delta \delta_{\sigma\sigma} \J &=& \displaystyle
  \dfrac{1}{\rho} \, \int_{\Omega} \mbox{div} \, \delta \Bsigma \cdot \Delta \mbox{div} \, \Bsigma  \, dV
+ \dfrac{1}{\rho \nu} \, \int_{\Omega} \delta \Bsigma \cdot \Delta \Bsigma \, dV \\
\Delta \delta_{pp} \J &=& \displaystyle
  \dfrac{1}{\rho \nu} \, \int_{\Omega} \delta p \bone  \cdot \Delta p \bone \, dV \\
\Delta \delta_{v\sigma} \J &=& \displaystyle
- \dfrac{1}{\rho \nu} \, \int_{\Omega} 2 \rho \nu \nabla^s \, \delta \bv 
 \cdot \Delta \Bsigma
- \dfrac{1}{\rho} \,\int_{\Omega} \left(\rho \nabla \delta \bv\, \bv  + \rho \nabla \bv \, \delta \bv  \right)  
 \cdot \mbox{div} \, \Delta \Bsigma  \, dV  \\
\Delta \delta_{vp} \J &=& \displaystyle
- \dfrac{1}{\rho \nu} \, \int_{\Omega} 2 \rho \nu \nabla^s \, \delta \bv 
 \cdot  \Delta p {\bf 1} \, dV \\
\Delta \delta_{\sigma p} \J &=& \displaystyle
  \dfrac{1}{\rho \nu} \, \int_{\Omega} \delta \Bsigma 
  \cdot \Delta p {\bf 1}  \, dV  \; .
\end{array}
\label{second}
\ee
The quantities 
$\Delta \delta_{\sigma v} \J$, 
$\Delta \delta_{pv} \J$ and
$\Delta \delta_{p \sigma} \J$
could be obtained by interchanging $\delta$ with $\Delta$.
Alternatively, the Newton tangent can also be computed by a standard difference
quotient procedure. The discretization of the domain is realized by triangles
equipped with different conforming interpolation schemes for the unknown variables.
For the velocities and pressure we use the finite element spaces
\eb
\displaystyle
\bV^h_k = \left\{ \bv \in H^1(\hat\Omega)^2 : \bv |_{\hat\Omega} \in P_k(\hat\Omega)^2 \; \forall \; \hat\Omega \right\} \subset \bV 
\ee
and
\eb
\displaystyle
\bQ^h_l = \left\{   p \in L_2(\hat\Omega)^2 :  p |_{\hat\Omega} \in P_l(\hat\Omega)^2 \; \forall \; \hat\Omega \right\} \subseteq \bQ \; ,
\ee
where the indices $k$ and $l$ denote the polynomial order of the standard
Lagrange polynomials $P_k(\hat\Omega)^2$ and $P_l(\hat\Omega)^2$, and
$\hat\Omega$ is the disretized version of the domain $\Omega$. For the stresses
a Raviart-Thomas finite element space, see e.g.\ \cite{RavTho:1977:amf} or
\cite{BreFor:1991:mah}, is applied in order to complete the $RT_mP_kP_l$ mixed
finite element
\eb
\bX^h_{m} = \left\{ \Bsigma \in H(\mbox{div},\hat\Omega)^2 : 
\Bsigma |_{\hat\Omega} \in RT_{m}(\hat\Omega)^2 \; \forall \; \hat\Omega \right\}  \subset \bX \; .
\ee
Again the index $m$ denotes the polynomial order of the vector-valued
Raviart-Thomas functions $RT_{m}$, which interpolate each row of the stress
tensor.  These functions are related to the edges of the triangles and their
normal components are continuous between two elements. In this work we use
the interpolation combination $RT_1P_3P_1$, which seems to be a sufficient
choice. For details about the implementation and other interpolation combinations
the reader is referred to \cite{SchNicSerNisOuaSchTur:2016:acs}.  
Finally, we obtain the system of equations for a typically element
\eb
\left[
\begin{array}{ccc}
\bk^e_{vv}       & \bk^e_{v\sigma}      & \bk^e_{vp}       \\
\bk^e_{\sigma v} & \bk^e_{\sigma\sigma} & \bk^e_{\sigma p} \\
\bk^e_{p v}      & \bk^e_{p \sigma}     & \bk^e_{p p}      \\
\end{array}
\right]
\left[
\begin{array}{c}
\Delta \bv^e \\
\Delta \Bsigma^e \\
\Delta \bp^e \\
\end{array}
\right]
=%\renewcommand{\arraystretch}{1.3}
\left[
\begin{array}{c}
\br^e_{v} \\
\br^e_{\sigma} \\
\br^e_{p} \\
\end{array}
\right] \; ,
\ee
where the entries of $\br^e$ and  $\bk^e$ result from (\ref{first}) and (\ref{second}).

\subsection{LSFEM with data (LSFEM+)}
The Navier-Stokes equations were formulated as a {\it minimization} problem in the previous
section.  This form suggests an simple adjustment to incorperate
experimental data.  Consider data $\bd \in\mathbb{R}^M$ which is approximated by
$Hw$ where $H:V\rightarrow\mathbb{R}^M$ is the {\it observation operator}.  The
implication is that $\bd \approx Hw$.  Then a possible LSFEM functional
$\mathcal{J}(w)$ becomes
\eb
\mathcal{J}^+(w) := \mathcal{J}(w) + \frac{1}{2}\sum_{i=1}^M \omega_i(d_i - H_iw)^2,
\label{eq:funaug}
\ee
i.e.\ differences between data and state are penalized.  Here $\omega_i$ are
per-measurement weights.

This is a useful modification in the following circumstances: a fluid-dynamics
experiment is performed, in which some (limited) quantities are measured.
E.g.~velocity $\bv$ at a limited number of spatial locations in $\Omega$;
pressure $p$ at a number of pressure taps on the surface $\Gamma$; or a force in
the form of an integral of $p$ over $\Gamma$.  These measurements may not
contain the Quantity-of-Interest (QoI) of the experimentor because it is
inaccessable or unmeasurable; or they may be noisy or of low-resolution.  It is
desired to estimate the QoI from the measured data, using knowledge of the
equations of motion of the fluid (in the form of a PDE solver).  However,
conditions in experiments are never precisely known: e.g.\ boundary conditions
are uncertain, and therefore can not be set in the solver.

By augmenting with data, the simulation becomes a better reproduction of the
experiment from which the data originated -- and the QoI can be estimated based
on a composite of knowledge derived from measurement and simulation.  The solver
can be thought of as performing PDE-based regression in the state-space.

In Section~\ref{s:bayes} we will formulate more precisely the problem that is
being solved, and justify the choice of the functional \eqref{eq:funaug} in
terms of Bayesian statistics.  A close connection to standard data-assimilation
techniques will be demonstrated.

%%%%%%%%%%%%%%%%%%%%%%%%%%%%%%%%%%%%%%%%%%%%%%%%%%%%%%%%%%%%%%%%%%%%%%%%%%%%%%%%
\section{Bayesian derivation of LSFEM data-assimilation}
\label{s:bayes}
In the following we formulate the data-assimilation problem, and provide a
Bayesian resolution.  The maximum {\it a posteriori} (MAP) estimate of the posterior is
shown to be exactly the solution of the augmented LSFEM problem of the previous
section.  The close connection to the Kalman filter (and equivalently 3DVAR with
a suitable background field) is demonstrated.  This development is made in
detail for the linear case.  The idea could be extended to the nonlinear case --
for an empirical evaluation the approach for N-S see
Section~\ref{s_fuckedifiknow}.  In conclusion, augmented LSFEM is shown to
produce a solution that is identical to standard data-assimilation approaches --
but for a substancially reduced computational cost.

\subsection{Problem statement and data model}
\begin{problem}[Data assimilation]
Let $w^\ast\in V$ be the unknown ground truth.  Knowledge of $w^\ast$ comes from
two sources:
\begin{enumerate}
\item Finite, noisy measurements $\bd\in\mathbb{R}^M$, obtained from $w^\ast$
  using the known observation operator $H:V\rightarrow\mathbb{R}^M$, i.e.\
\eb
\bd \simeq Hw^\ast.
\label{eq:stat1}
\ee
\item Approximate knowledge of the governing equations and boundary-conditions,
  i.e.\
\eb
 Lw^\ast-f\simeq 0\:\:\text{on $\Omega$}, \qquad Bw^\ast-g\simeq 0\:\:\text{on $\Gamma$}.
\label{eq:stat2}
\ee
\end{enumerate}
Estimate $w^\ast$.
\end{problem}

To characterize the inexact equalities in \eqref{eq:stat1}--\eqref{eq:stat2}, we
introduce statistical models.  In \eqref{eq:stat1} assume that observations are
corrupted by unbiased Gaussian measument noise with known symmetric
positive-definite covariance matrix $\bR\in\mathbb{R}^{M\times M}$, i.e.\
\eb
\bd = H w^\ast + \Bvarepsilon, \quad \Bvarepsilon\sim\mathcal{N}(\bzero, \bR).
\label{eq:model12}
\ee
The assumption of unbiased noise is reasonable if all {\it known} sources of
bias are corrected for.  (Approximately) known covariance is a reasonable
assumption if experimental noise has been estimated concurrent with the
measurements.  Both these are features of modern experimental methodologies.  No
stochastic term for ``model inadequacy'' in $H$ \cite{KenOha2001} is included,
under the assumption that $H$ is a relatively simple operator that merely
extracts necessary quantities from the state $w^\ast$, and does not contain
any physical modelling or discretization.  This is the case for many observation
operators of practical interest, including the ``point velocity observations''
used in the following section.

\subsection{Model inadequacy in the governing equations}
In contrast \eqref{eq:stat2} contains both physical modelling, discretization,
and potentially unknown/uncertain equations-of-state and boundary-conditions.
All these are reasons for the ground truth to not exactly satisfy the (discretized)
governing equations \eqref{eq:varpart}.  In variational form consider a modified
version of \eqref{eq:varpart} which the ground-truth does satisfy:
\eb
Q(w^\ast, v) = l(v) + \delta(v) \quad \forall v\in V,
\label{eq:blah}
\ee
where the new functional $\delta: V\rightarrow \mathbb{R}$, represents the
discrepancy, and could be defined by $\delta(v) := l(v) - Q(w^\ast, v)$ if
the ground truth $w^\ast$ were known.  In practice $\delta(v)$ must be modelled, and
following~\cite{NguPer2016} we treat it as a Gaussian process
\eb
\delta(v) \sim \mathcal{GP}(0, k(w,v)),
\label{eq:prior1}
\ee
with zero mean and covariance operator $k:V\times V\rightarrow \mathbb{R}$.
The resulting equation for $w$ is a stochastic PDE:
\eb
Q(w, v) = l(v) + \delta(v) \quad \forall v\in V.
\label{eq:blah15}
\ee
In~\cite{NguPer2016} the authors consider a parameterized family of covariance
operators $k(\cdot,\cdot)$:
\[
k(w,v;\Btheta) := \theta_1 (w,v)_\Omega + \theta_2 (\nabla w,\nabla v)_\Omega,
\]
representing the $H^1(\Omega)$ scalar product.  Here we propose a covariance
operator as a scalar product informed by the governing PDE:
\eb
k(w,v;\Btheta) := \theta_1 (Lw,Lv)_\Omega + \theta_2 (Bw,Bv)_\Gamma,
\label{eq:blah2}
\ee
which is symmetric positive-definite 
\[
k(w,v)=k(v,w),\quad k(v,v)>0\:\:\text{for $v\neq 0$},\quad k(v,v)=0\:\:\text{for $v=0$}
\]
by construction, provided $L$ is bounded below.  This condition is in any case a
prerequisite for the stability and accuracy of LSFEM -- in particular it is
true for the Navier-Stokes operator of the previous section.

\subsection{Calibration}
It remains to compute the posterior $w\mid\bd$ under the statistical model
\eqref{eq:model12} and prior \eqref{eq:prior1} -- i.e.\ the updated estimate of
the full state.  First it is necessary to discretize to obtain
finite-dimensional equations, with basis $(\Bphi_1,\dots,\Bphi_N)$:
\[
w \simeq w_h := \sum_{i=1}^N \bw_i \Bphi_i,
\]
where discrete quantities are represented by bold type.  Given this
\eqref{eq:model12} becomes
\[
\bd = \bH\bw + \Bvarepsilon,\quad \bH_{ij}:= (H\Bphi_j)_i,
\]
and the prior on the disretized discrepancy $\Bdelta$ is
\[
\Bdelta_0 \sim \mathcal{N}(\bzero, \BSigma), \quad \BSigma_{ij} := k(\Bphi_i,\Bphi_j).
\]
Furthermore, from \eqref{eq:blah15}
\eb
\bQ\bw = \bl + \Bdelta, \quad \bQ_{ij} := Q(\Bphi_i,\Bphi_j),\:\: \bl_{j} := l(\phi_j).
\label{eq:blah3}
\ee
Note that, if the weights in the definitions of $Q(\cdot,\cdot)$ and
$k(\cdot,\cdot)$ are equal, then $\bQ\equiv\BSigma$, which is a direct
consequence of the particular choice of prior covariance for $\Bdelta$.  This
observation, together with \eqref{eq:blah3} implies a prior on $\bw$:
\[
\bw_0 \sim \mathcal{N}( \bQ^{-1} \bl, \bQ^{-1}\BSigma\bQ^{-1}) = \mathcal{N}( \bw^\circ, \bQ^{-1}),
\]
where $\bQ \bw^\circ = \bl$ defines $\bw^\circ$ as the solution of the original
LSFEM discretization (without a data term).  Having eliminated $\Bdelta$ we can
apply Bayes' theorem to $\bw$:
\[
 \rho(\bw \mid \bd) \propto \rho(\bd \mid \bw) \cdot \rho_0(\bw).
\]
Since all stochastic terms are Gaussian the posterior probability density function (pdf) is
\begin{align*}
\rho(\bw \mid \bd) &= \rho_\varepsilon(\Bvarepsilon) \cdot \rho_0(\bw) \\
                   &= \exp\left[-\frac{1}{2} \Bvarepsilon^T \bR^{-1}\Bvarepsilon -\frac{1}{2}(\bw-\bw^\circ)^T\bQ(\bw-\bw^\circ)\right] \, . 
\end{align*}
The MAP estimate of $\bw$ is therefore the solution of the optimization problem
\[
\bw^+ := \mathrm{argmin}_\bw \;
\left[\frac{1}{2}(\bd - \bH\bw)^T \bR^{-1}(\bd - \bH\bw) + \frac{1}{2}(\bw-\bw^\circ)^T\bQ(\bw-\bw^\circ)\right],
\]
and the relation to LSFEM is immediately obvious.  In particular, setting
1st-variations of the cost-function to zero gives
\begin{align}
   \bw^T\bQ\delta\bw + \bw^T \bH^T \bR^{-1} \bH \delta\bw &= \bd^T \bH^T\bR^{-1}\delta \bw + \bl^T \delta\bw, \quad \forall\,\delta\bw\in\bV
   \label{eq:blah4}
\end{align}
which is exactly the discrete version of the LSFEM problem with data-term from
\eqref{eq:funaug}: Find $w\in V$ such that 
\eb 
\underbrace{(Lw, L\delta
  w)_\Omega + (Bw, B\delta w)_\Gamma}_{Q(w,\delta w)} + \underbrace{(Hw, H\delta
  w)_R}_{T(w,\delta w)} = \underbrace{(f,L\delta w)_\Omega + (g,B\delta
  w)_\Gamma}_{l(\delta w)} + \underbrace{(\bd,H\delta w)_R}_{t(\delta w)} \quad
\forall \delta w\in V.
\label{eq:varfull}
\ee 
provided that $[\bR]_{ii} = \omega_i$ and $[\bR]_{ij}=0$ for $i\neq j$.  In
the above, the inner-product for the data is defined as
\[
(\bx, \by)_R := \bx^T \bR^{-1} \by.
\]
Thus the Bayesian framework -- with appropriate statistical modelling -- leads
us to the LSFEM+.

Note that the connection between LSFEM+ and the statistical data-assimilation
problem relies upon the particular choice of an {\it additive} discrepency
$\Bdelta$ in \eqref{eq:blah}, with the specific covariance defined by
\eqref{eq:blah2}.  If an alternative covariance structure for $\Bdelta$ is
chosen, e.g.\ from \cite{NguPer2016}, LSFEM does not provide a solution of the
corresponding inverse problem.

\subsection{Connection to the Kalman filter}
We would like to describe the the solution of \eqref{eq:varfull} (denoted
$w^+$), in terms of the solution to the same equation without data
\eqref{eq:varpart} ($w^\circ$).  The goal is to show the result is identical to the
application of a Kalman filter.  

Starting from \eqref{eq:blah4}, cancelling $\bl$ using $\bQ\bw^\circ=\bl$, and
rearranging gives
\begin{align}
\bw^+ &= \bw^\circ + (\bQ + \bH^T\bR^{-1}\bH)^{-1} (\bH^T\bR^{-1}\bd - \bH^T\bR^{-1}\bH\bu^\circ) \\
      &= \bw^\circ + \left[ (\bQ + \bH^T\bR^{-1}\bH)^{-1} \bH^T\bR^{-1}\right] (\bd - \bH\bu^\circ) \\
      &= \bw^\circ + \underbrace{\bQ^{-1} \bH^T (\bH\bQ^{-1}\bH^T + \bR)^{-1}}_{\bK} (\bd - \bH\bu^\circ),
\end{align}
where the final equality is obtained using a modified version of the matrix identity % Woodbury
\[
(\bQ + \bH^T \bR^{-1} \bH)^{-1} \bH^T \bR^{-1} \equiv \bQ^{-1} \bH^T (\bH\bQ^{-1}\bH^T + \bR)^{-1}.
\]
The matrix $\bK$ is the well-known optimal Kalman gain, under the statistical
model \eqref{eq:model12}, and the background covariance model $\bw \sim
\mathcal{N}(\bw^\circ, \bQ^{-1})$.  In fact $\bw^+$ is the MAP estimate of
$\bw|\bd$ with distribution
\[
\bw|\bd \sim \mathcal{N}\left(\bw^+, (\bI - \bK \bH)\bQ^{-1}\right).
\]
This result thereby also provides an expression for the posterior covariance of the
full state $\bw$.

%%%%%%%%%%%%%%%%%%%%%%%%%%%%%%%%%%%%%%%%%%%%%%%%%%%%%%%%%%%%%%%%%%%%%%%%%%%%%%%%
\section{Application: Laminar Flow over a Backward-Facing Step}
\label{s_fuckedifiknow}
The phenomena of flow separation and subsequent re-attachment are important in
many engineering situations.  For example in pipe flow separation is undesirable
and leads to reduction in flow rate for a given pressure-drop, requiring
additional pumps or fans.  In other circumstances separation may be desirable,
to promote mixing or flame stabilization in combustion.  However the presence
and location of separation on curved surfaces, and the location of re-attachment
in general, are difficult to predict, and are sensitive to small
uncertainties in inflow conditions and fluid properties.

We aim to show that using down-stream measurements and simulation it is possible
to accurately determine the location of the re-attachment point in the presence
of uncertain boundary-conditions.  We consider a backward-facing step in laminar
flow for which accurate experimental velocity profiles are available, Denham and
Patrick~\cite{DenPat1974,BarFon2002}.  This experiment is especially appropriate
for this study, as it is supposed to model an ideal situation in which the
inflow-boundary is a fully-developed channel-flow (with a parallel, parabolic
velocity-profile), but due to experimental limitations the inflow is far from
fully-developed.  In~\cite{DenPat1974} the actual inflow profile is also
measured, and this will be used as a reference.

\begin{figure}[h!] \unitlength 1 cm
\begin{picture}(14,6.5)
\put(0.5,0.1){\scalebox{0.90}{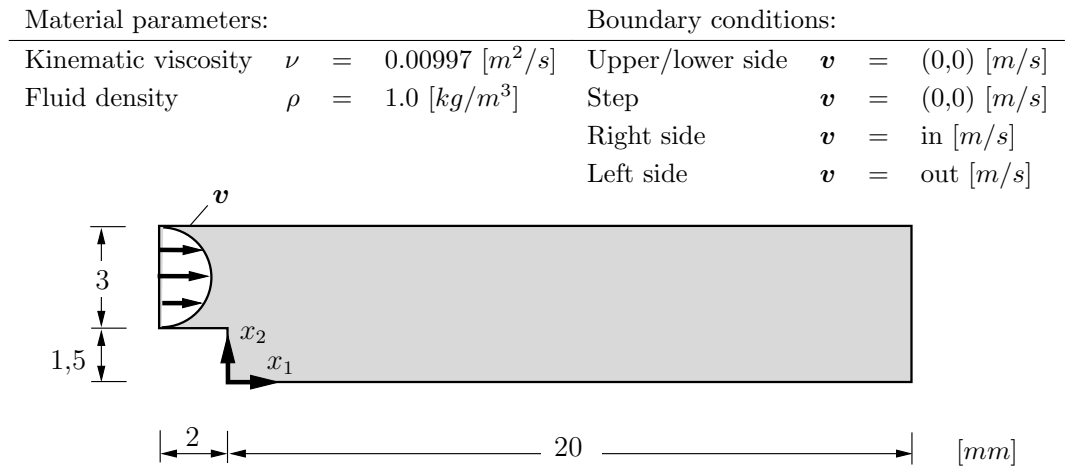}}
\put(12.5,0.2){$[mm]$}
\put(-0.2,4.9){
\renewcommand{\arraystretch}{1.2}
\begin{tabular}{lrcl}
\mbox{Material parameters}: \hspace{-2cm} \phantom{1} &&&\\
\hline
\mbox{Kinematic viscosity} & $\nu$ &=& 0.00997 $[m^2 / s]$\\
\mbox{Fluid density} & $\rho$ &=& 1.0 $[kg / m^3]$ \\
&&&\\
&&&
\end{tabular}
}
\put(7.2,4.9){
\renewcommand{\arraystretch}{1.2}
\begin{tabular}{lrcl}
\mbox{Boundary conditions}: \hspace{-2cm} \phantom{1}  &&&\\
\hline
\mbox{Upper/lower side} & $\bv$ &=& (0,0) $[m / s]$\\
\mbox{Step} & $\bv$ &=& (0,0) $[m / s]$\\
\mbox{Right side } & $\bv$ &=& in $[m / s]$\\
\mbox{Left side } & $\bv$ &=&  out $[m / s]$
\end{tabular}
}
\end{picture}
\caption{Material parameters and geometry of the backward-facing step.}
\label{setup3}
\end{figure}

The setup and material parameters of the backward-facing step can be found in
Figure~\ref{setup3}.  The Reynolds number is $\mathrm{Re}=191$ based on the step
height and inflow peak velocity~\cite{DenPat1974}.  The circulation length is
roughly $x_l = 12.7$.  We prescribe no-slip BCs on the lower- and upper walls
and a zero pressure BC on the right-hand outflow.  On the left-hand side we
consider a variety of BCs representing different states of knowledge of the
inflow condition.

Due to the low Reynolds-number the flow is laminar, and therefore no turbulence
modelling is required, eliminating a potential source of modelling inadequacy
which would have been difficult to estimate and control~\cite{KenOha2001}.
Errors in our solutions are therefore all due to discretization, and due to the
imprecisely known inflow boundary-condition.  We
examine the effect of adding data on these two error sources.  We numerically
investigate: (i) improving the accuracy of the predictions by adding data in
Section~\ref{s_example2}; (ii) solving the inverse problem in which the inflow
BC is completely unknown and unspecified~\ref{s_bcerror1}; and (iii) exploiting
partial, uncertain knowledge of the inflow to define the flow~\ref{s_bcerror2}.

The discretization uses cubic $RT_1P_3P_1$ elements, on two triangular meshes: a
coarse mesh with 528 nodes and a fine mesh of 8770 nodes.  Results will be
compared to experimental velocity profile measurements at four downstream
locations: $x=3$, $x=6$, $x=9$, $x=12$, and the inflow plane $x=-2$.

%%%%%%%%%%%%%%%%%%%
\subsection{Data Acquisition and Processing}
\label{s_data}
\begin{figure}[h!]
\includegraphics[width=\textwidth]{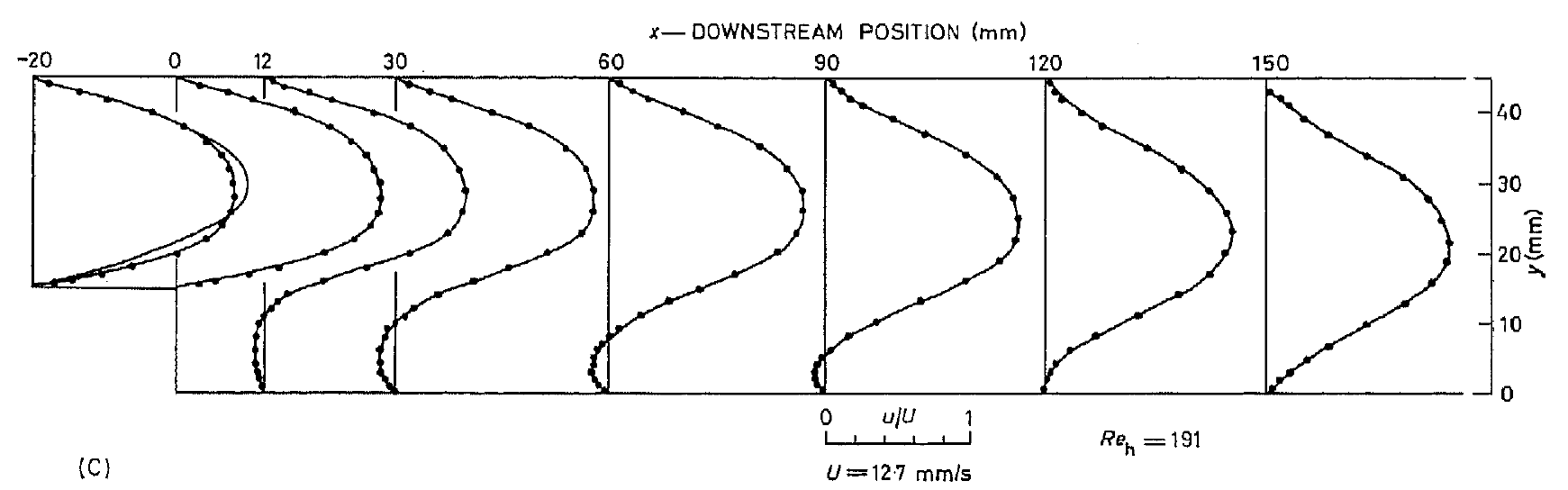}
\caption{Reproduction of Figure~5(c) from Denham and
  Patrick, 1974~\cite{DenPat1974}.  Velocity profiles for a backward-facing
  step at $\mathrm{Re}=191$.}
\label{f_scan}
\end{figure}

Denham and Patrick~\cite{DenPat1974} used a highly accurate non-invasive laser
D\"opplar anemometry system to measure the $x$-component of the flow velocity
vector.  They report smooth, well-resolved velocity profiles and give error
estimates.  The anemometry itself is described as being accurate to
$0.003\,\mathrm{mm/s} + 2\%$ of its value.  In addition we obtained the data
{\it in fine}, by electronically scanning the hand-plotted graphs
of~\cite{DenPat1974} (the only available source), see Figure~\ref{f_scan}.  An
additional inaccuracy arises in this process which we estimated to be $5\%$.
These values inform our choice of $\bR$.  Of other experimental errors, the
authors consider 3d effects to be negligible, but report the non-fully developed
inflow, and a time-varying flow-rate of 0.5\% as potentially significant error
sources.  These errors will be accounted for via the uncertain inflow condition.

\begin{figure}[h!]  
\centering
  \includegraphics[width=0.5\textwidth]{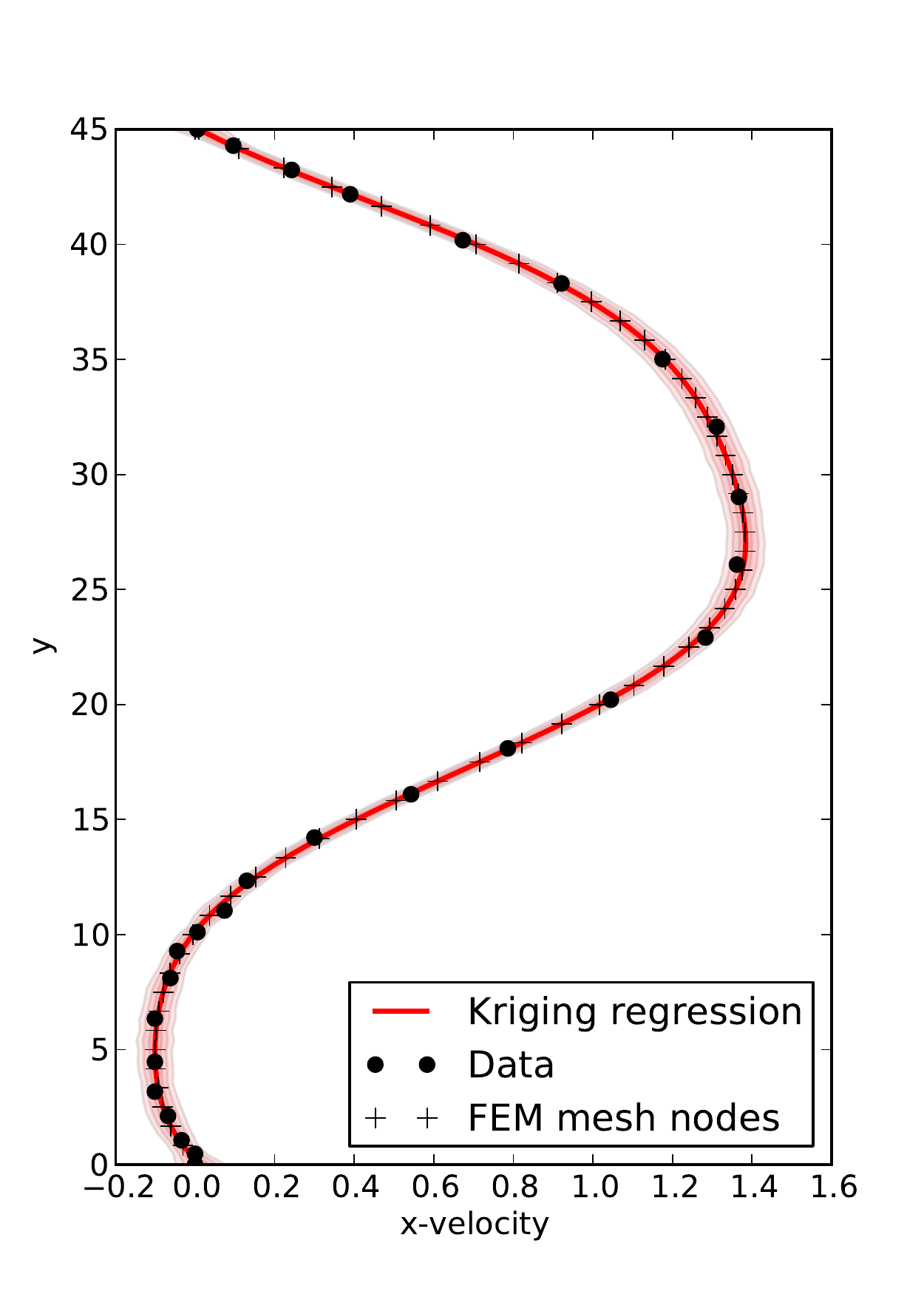}
\caption{Horizontal velocity component at $\mathrm{Re}=191$, $x=3.0$.
  Original measurements, Kriging response mean and 3$\sigma$ with
  assumption of $5\%$ noise in measurements, and values interpolated
  at computational nodes.}
\label{f_kriging}
\end{figure}

For our implementation it is convenient to have measurement data (and error)
available at the nodes of the finite-element mesh.  For this purpose we regress
the digitized velocity profile points with a Kriging
surface~\cite{DebDwiBij2011b}.  Kriging is a stochastic response surface
methodology, and as such can account for noise in data, and predict
uncertainties at reconstruction points \cite{Oha1992,WikBer2007}.  We use the
capability to construct $\bR$ at the FE-mesh nodes.  See Figure~\ref{f_kriging},
for an example reconstruction at $x=3.0$.

%%%%%%%%%%%%%%%%%%%%%%%%%%%%%%%%%%%%%%%%%%%%%%%%%%%%%%%%%%%%%%%%%%%%%%%%%%%%%%%%
\subsection{Error Study for the Simulation without Data}
\label{s_nodata}
Before looking at data assimilation we investigate the performance of the
numerical simulation in the absence of experimental data in order to study
sources of error.  Critically here we examine error with respect to experimental
data -- not with respect to a fine-grid simulation.  Therefore not only are
discretization errors visible, but also modelling inadequacy, and uncertainties
in the boundary conditions and material properties~\cite{KenOha2001}.  We
demonstrate that for a reasonably fine mesh, the inflow BC uncertainty dominates
the discretization error, and therefore is more important for correctly
reproducing the experiment, and presumably for prediction.

Firstly we consider the case of a parabolic velocity profile at the inflow
boundary (Poiseuille flow):
\eb
\bv = (4 v_{max} \, y (h - y) / h^2 , 0)^T
\ee
where $h \in [0,3]$ is the height of the inflow and the assumed peak inflow
velocity is $v_{max} = 1.81$.  In the absence of measurement data, this is the
simulation practitioner's best-guess of the inflow condition, representing
fully-developed laminar pipe-flow, see Section~\ref{s_data}.

\begin{figure}[ht]
\centering
\subfigure[$x=3$]{
\includegraphics[width=0.47\textwidth]{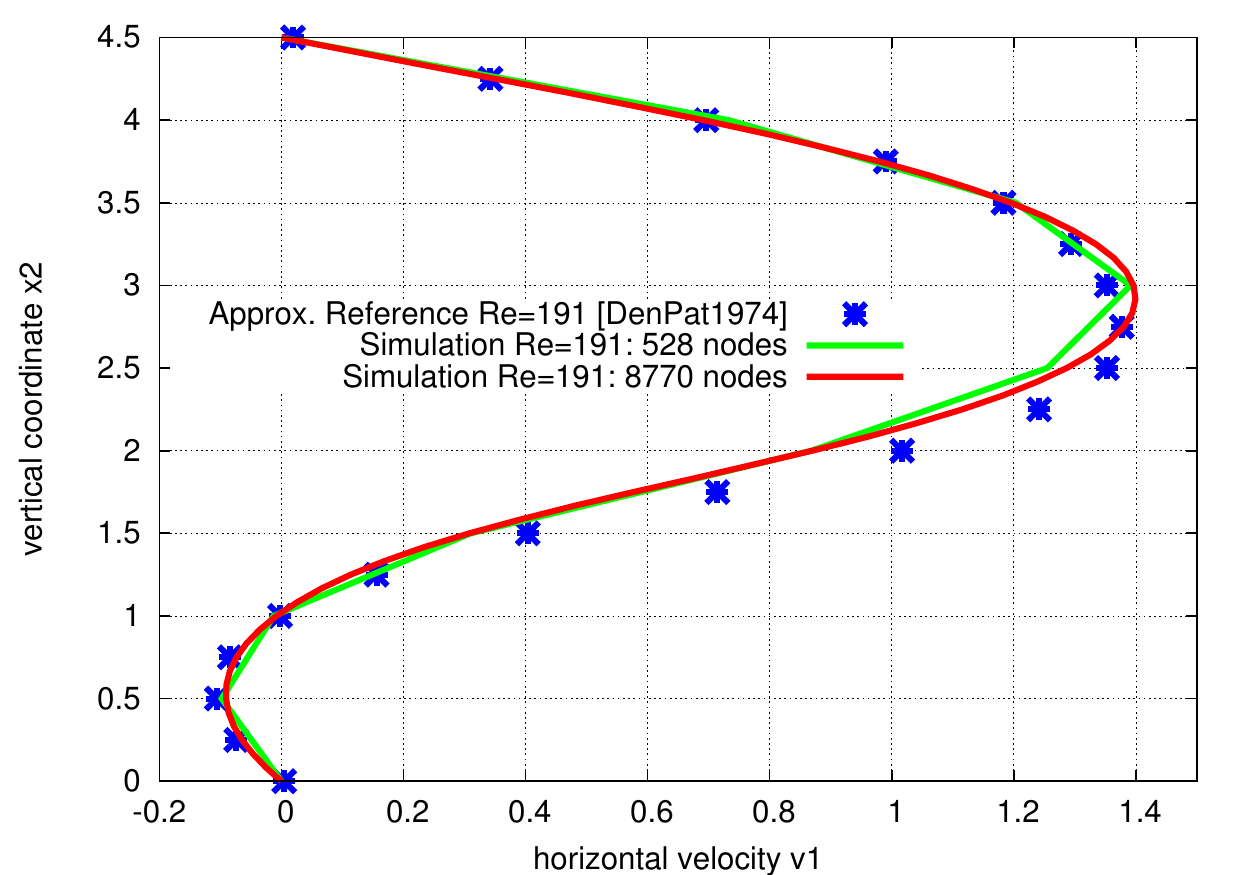}
}
\subfigure[$x=6$]{
\includegraphics[width=0.47\textwidth]{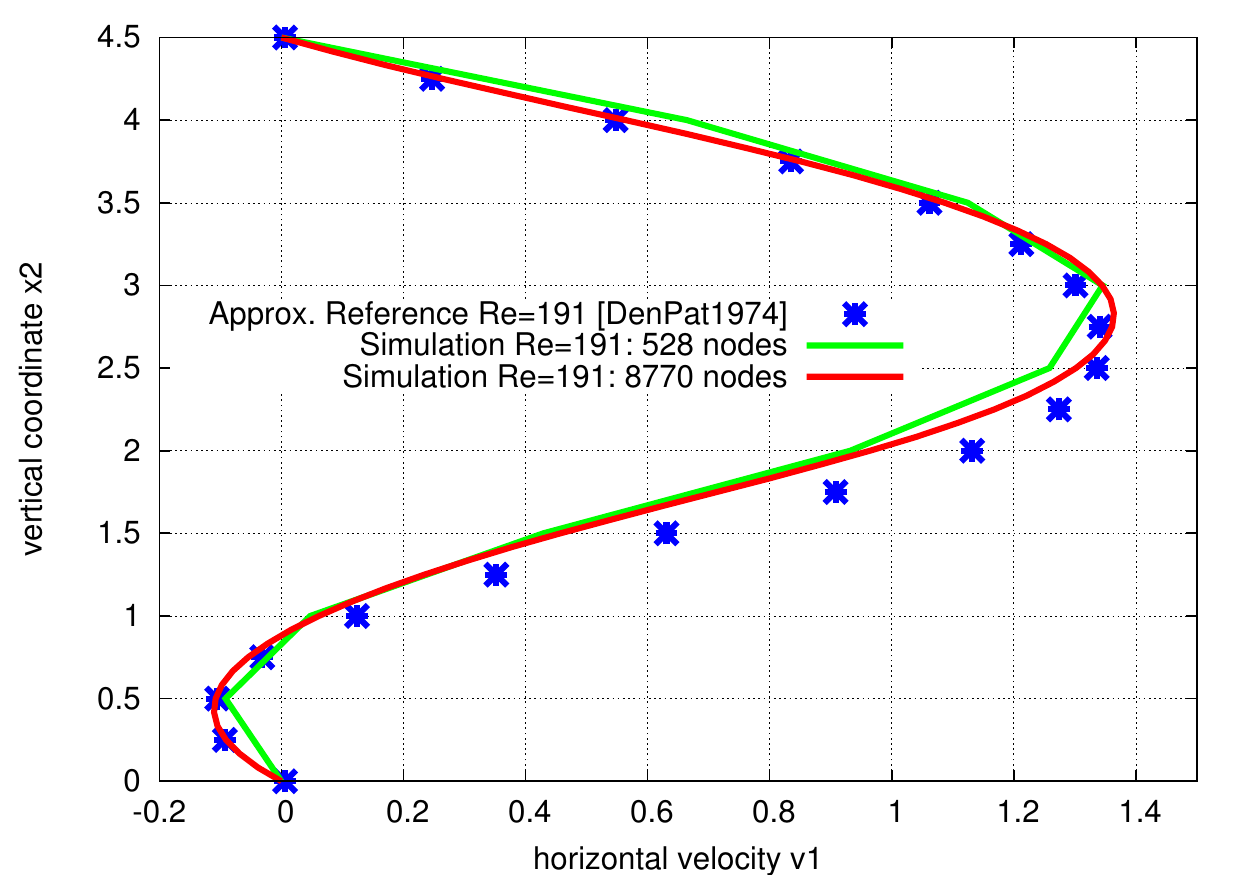}
}\\
\subfigure[$x=9$]{
\includegraphics[width=0.47\textwidth]{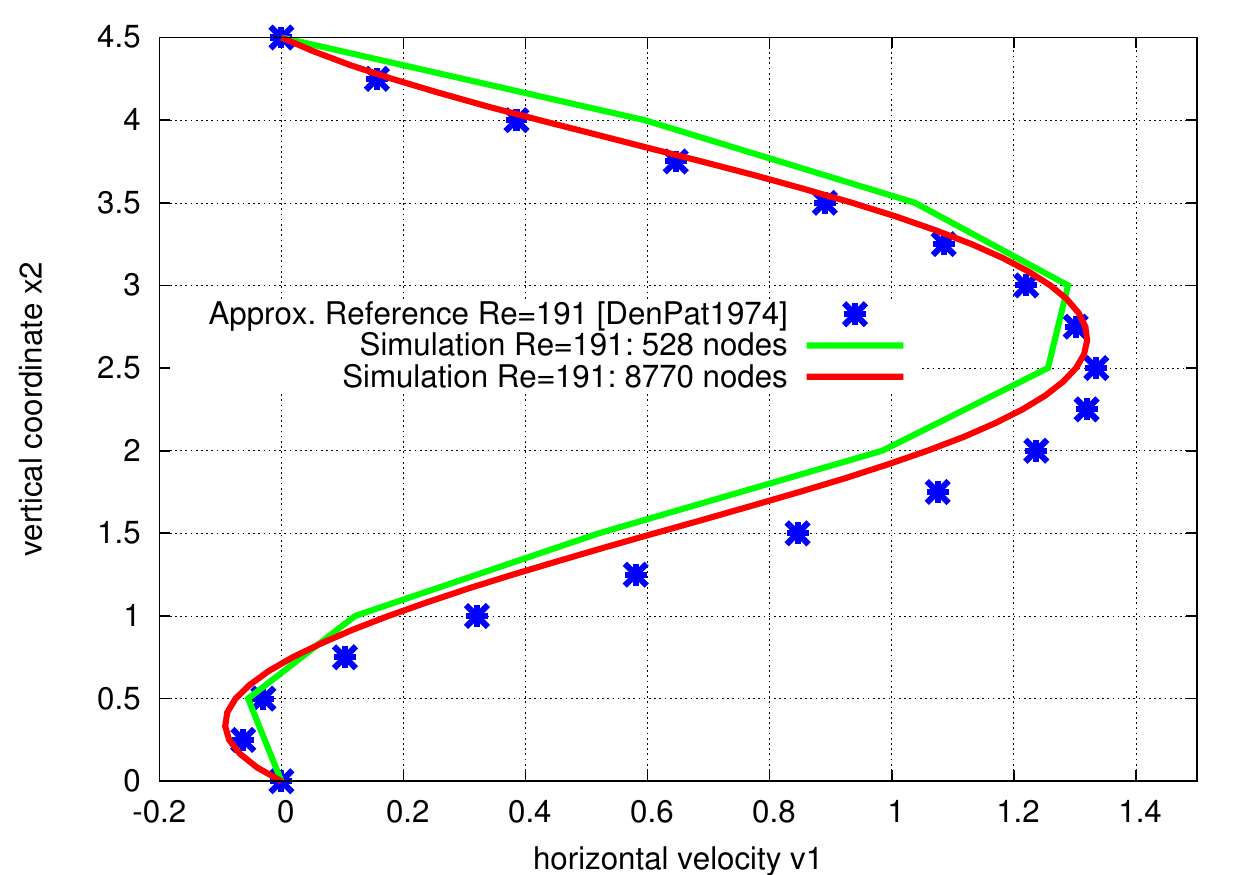}
}
\subfigure[$x=12$]{
\includegraphics[width=0.47\textwidth]{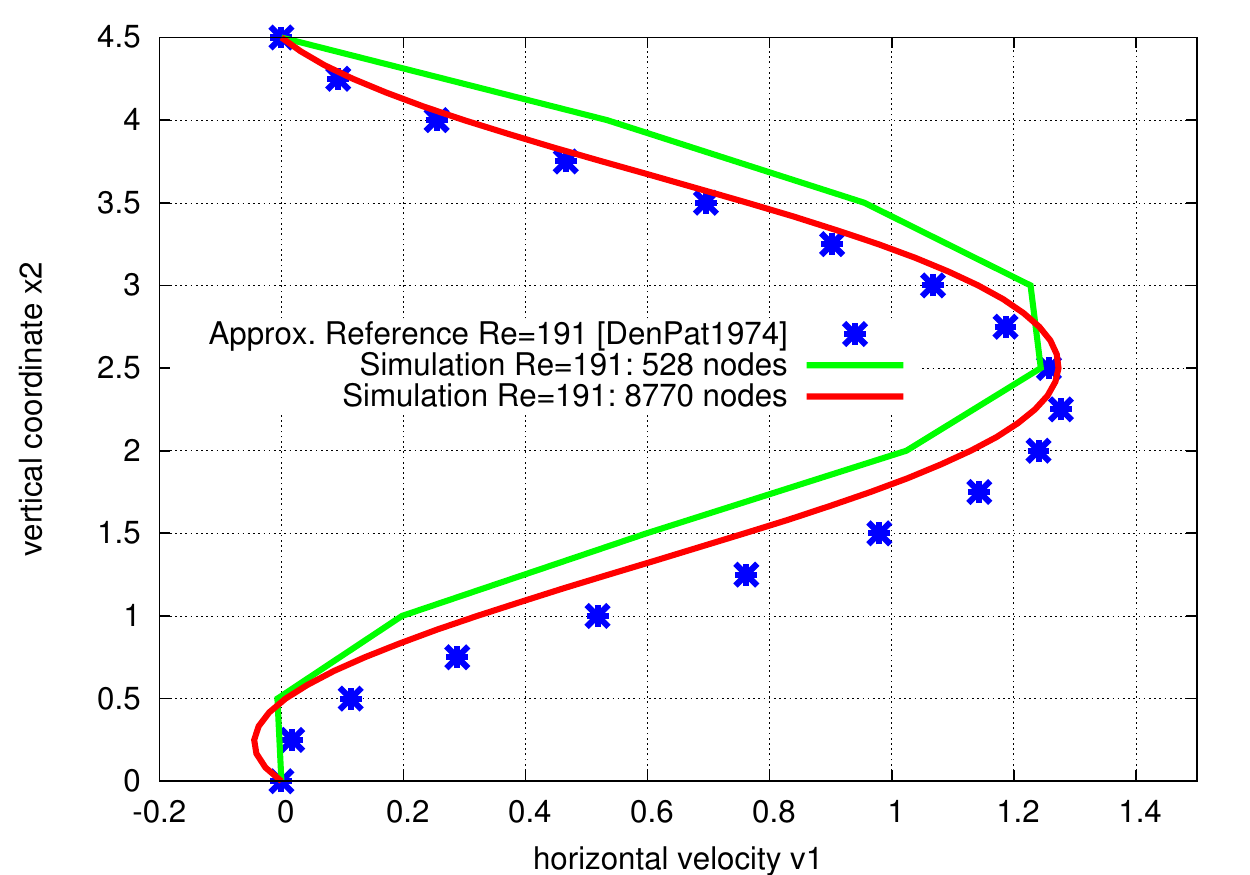}
}
\caption{Horizontal velocity profiles.  Parabolic inflow condition on a coarse and fine mesh.}
\label{velopara1}
\end{figure}

Figure~\ref{velopara1} shows results for the discretizations of 528 and 8770
nodes as well as the reference solution of~\cite{DenPat1974} at four
$x$-locations.  It can be seen that although the fine grid solution gives a
substantial accuracy improvement over the coarse grid, neither approaches the
experimental reference.  In particular the flow reattachment location in the
experiment is at $x_l=12.7$, while on the fine grid it is at $x_l=14.5$.  Further
refinement of the mesh shows no substantial change.  Given that modelling
inadequacy is also expected to be small, we conclude that the largest part of
the remaining error is due to imprecise reproduction of the experimental
conditions.

\begin{figure}[ht]
\centering
\subfigure[$x=3$]{
\includegraphics[width=0.47\textwidth]{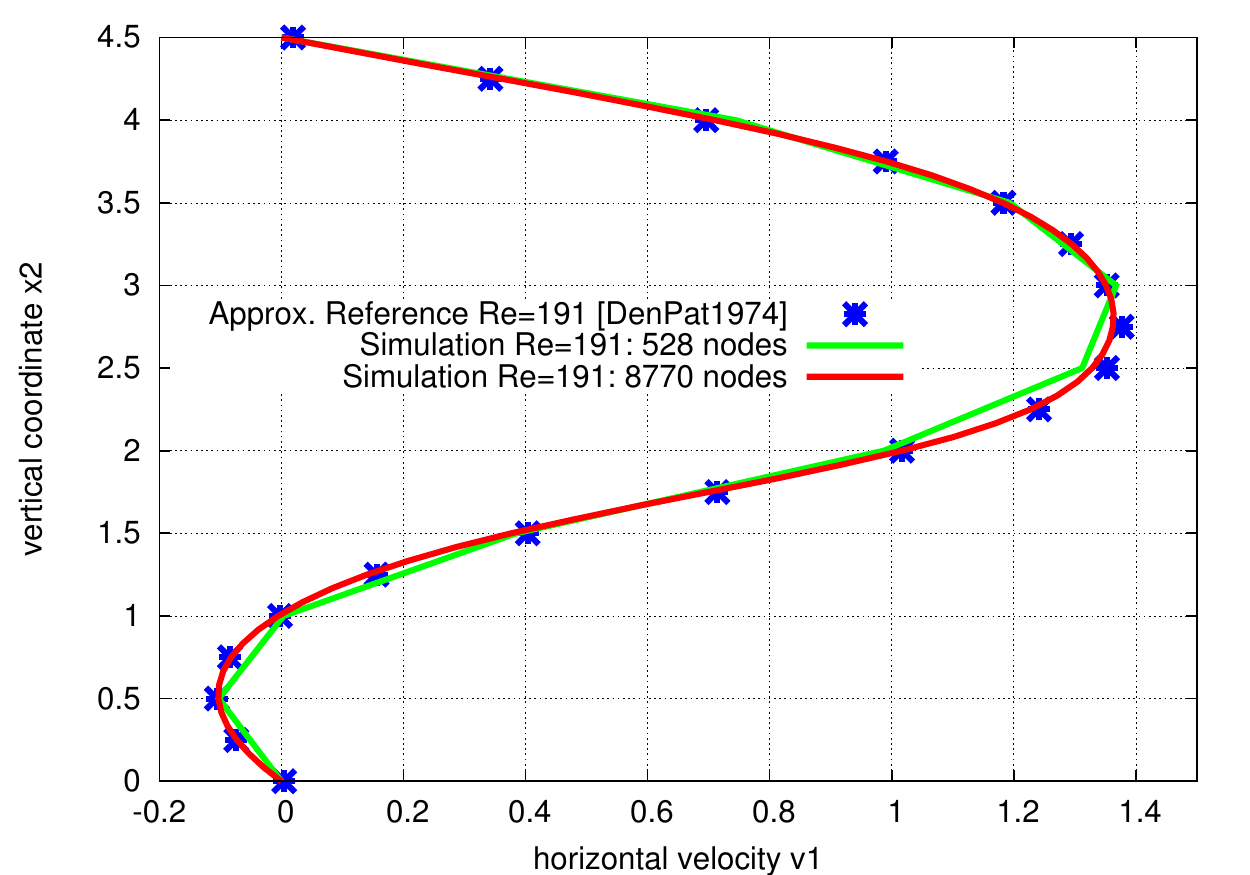}
}
\subfigure[$x=6$]{
\includegraphics[width=0.47\textwidth]{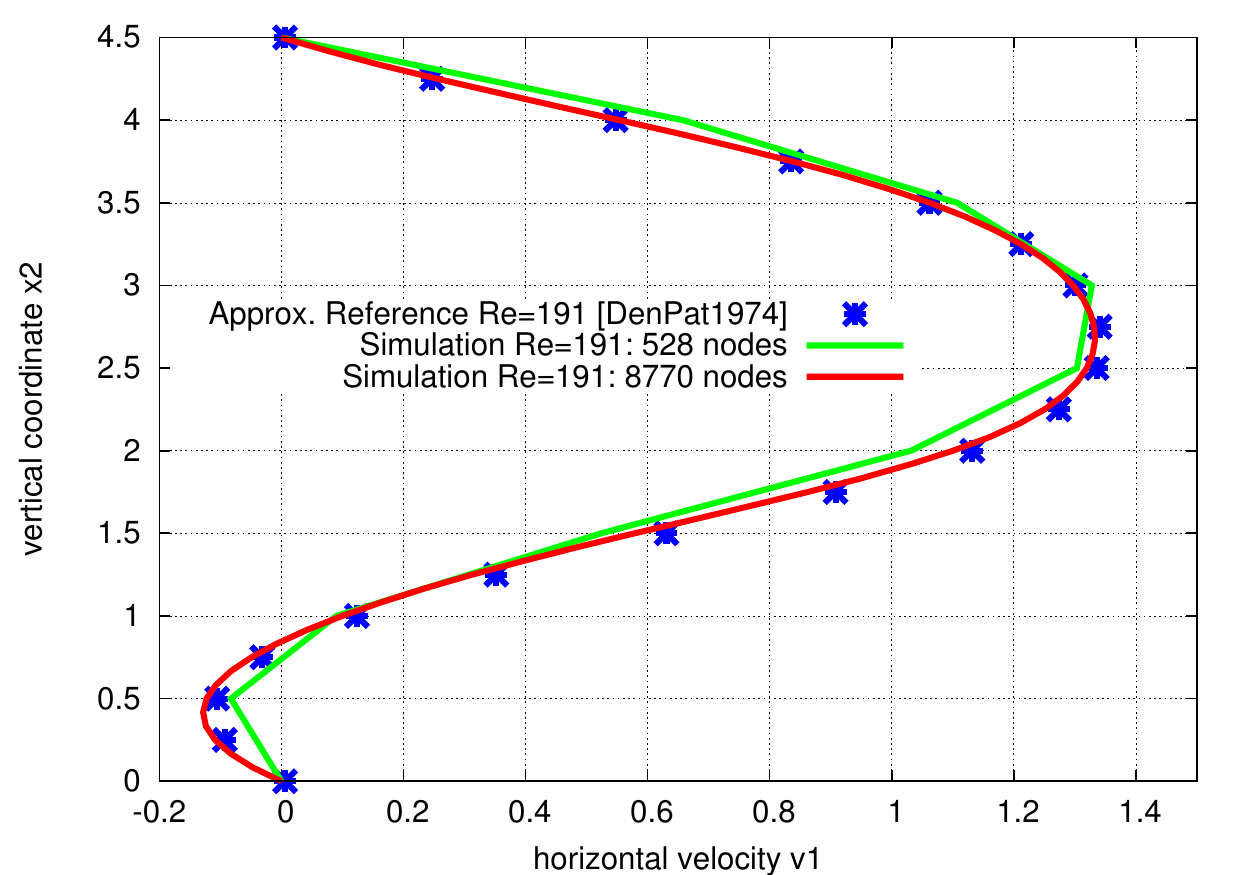}
}\\
\subfigure[$x=9$]{
\includegraphics[width=0.47\textwidth]{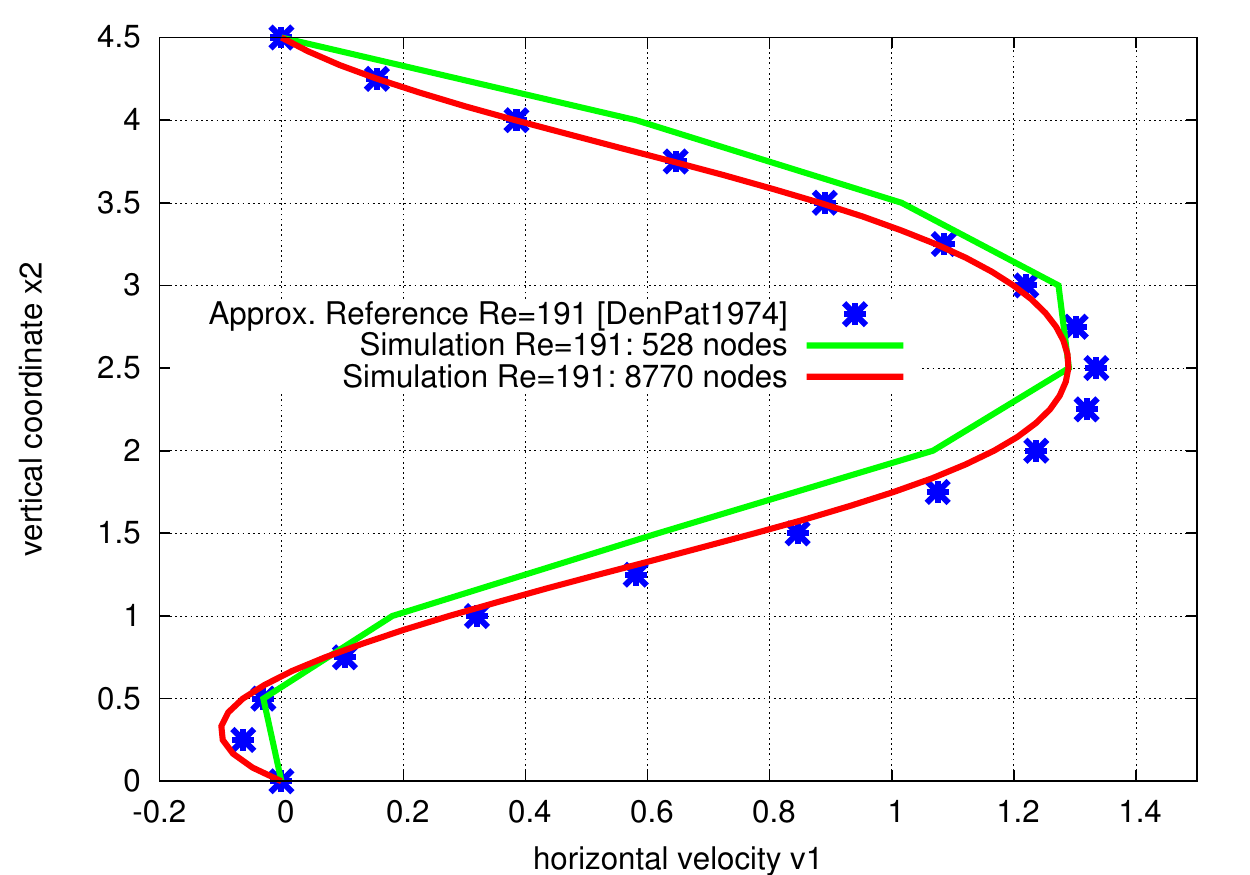}
}
\subfigure[$x=12$]{
\includegraphics[width=0.47\textwidth]{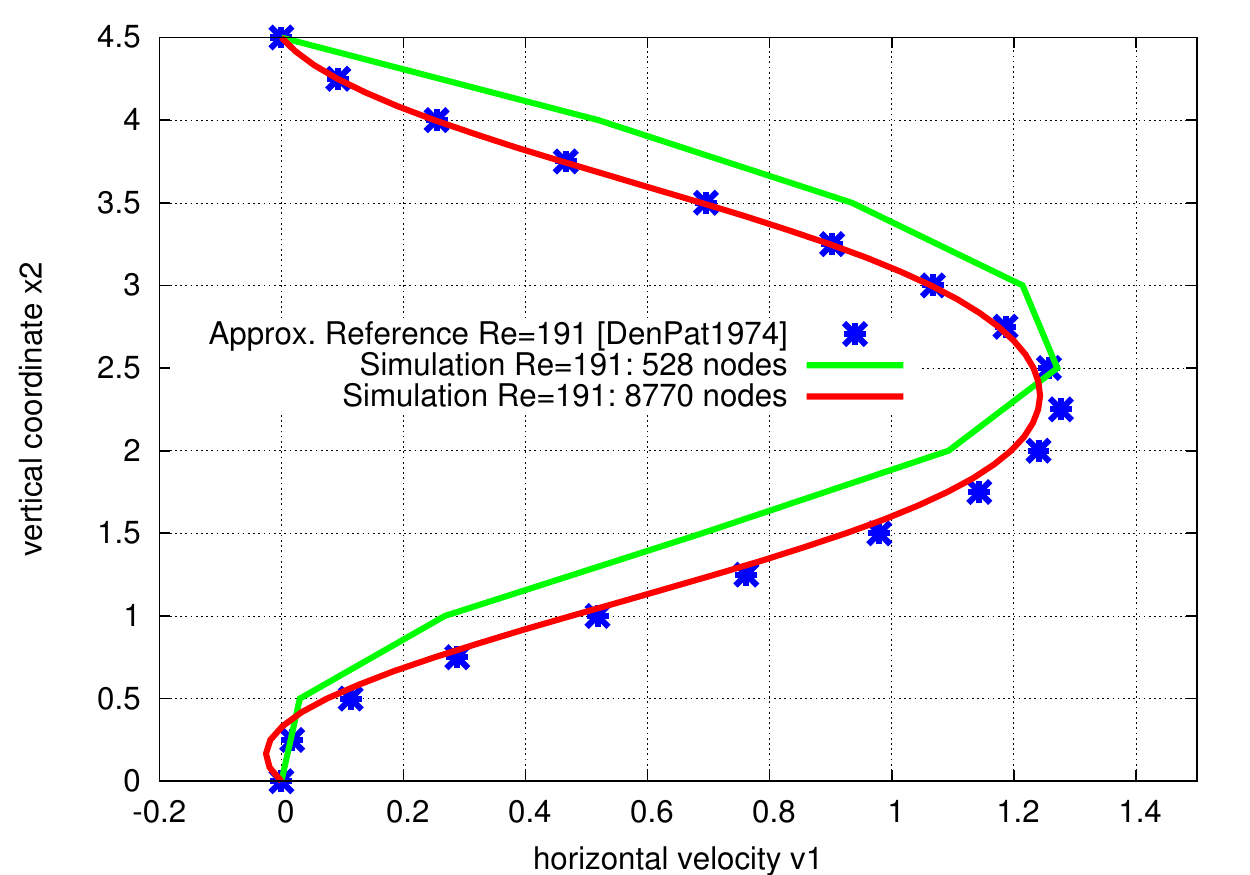}
}
\caption{Horizontal velocity profiles.  Inflow from experiment, coarse
  and fine grids.}
\label{veloexp1}
\end{figure}

From Figure~\ref{f_scan} we see that actual the inflow profile measured in the
experiment significantly deviates from a parabola.  Therefore in the following
computations the inflow BC is changed to the measured velocities from the
experiment, interpolated with Kriging.  The previous computations are repeated
with the new BC, and results depicted in Figure~\ref{veloexp1}.  It can be seen
that for the fine grid the largest part of the discrepancy is removed.  The
remaining small differences can be attributed to uncertainties in the unmeasured
vertical component of the inflow velocity.  The recirculation length is now
predicted as $x_l=13.3$, also significantly improved.  The coarse grid solution
remains poor --- discretization error clearly dominates here.

This demonstrates one case in which the leading-order error with respect to
reality can be discretization error or BC uncertainty, depending on the mesh.
We aim now to estimate and reduce both these types of error with the injection
of measurement data downstream of the inflow.

%%%%%%%%%%%%%%%%%%%%%%%%%%%%%%%%%%%%%%%%%%%%%%%%%%%%%%%%%%%%%%%%%%%%%%%%%%%%%%%%
\subsection{Correcting Discretization Error with Experimental Data}
\label{s_example2}
For the coarse grid solution in Figure~\ref{veloexp1} we observed a discretization error
which appeared to dominate the total error with respect to reality.  We examine
the ability of data injection downstream to reduce this large discretization
error.  Therefore, we study the same cases as previously, with the inflow BC
specified from the experiment.

\begin{figure}[h!] \unitlength 1 cm
\begin{picture}(14,4.6)
\put(0.0,0.0){\scalebox{0.5}{\includegraphics{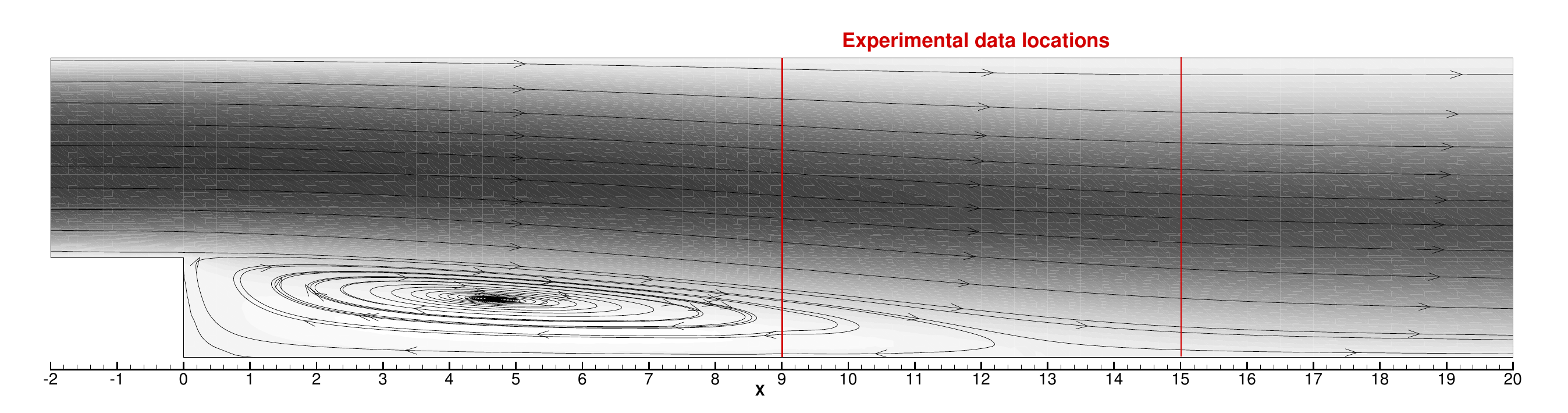}}}
\end{picture}
\caption{Streamlines and contours of horizontal-flow velocity for the
  simulation on the fine mesh, with experimental inflow BC, and
  experimental data inserted at $x=9$.  Alternative data location at
  $x=15$ also marked.}
\label{velofielddata}
\end{figure}

\begin{figure}[ht]
\centering
\subfigure[$x=3$]{
\includegraphics[width=0.47\textwidth]{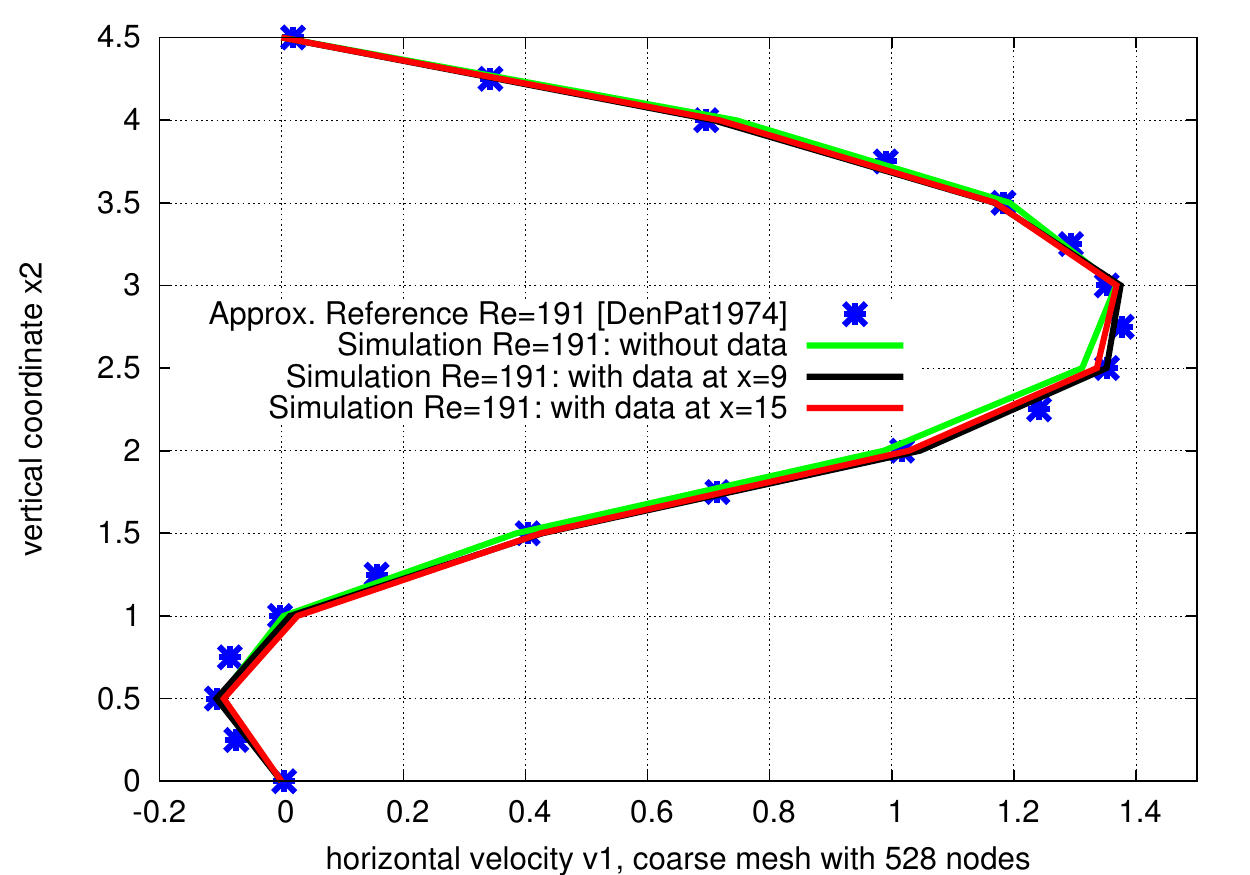}
}
\subfigure[$x=6$]{
\includegraphics[width=0.47\textwidth]{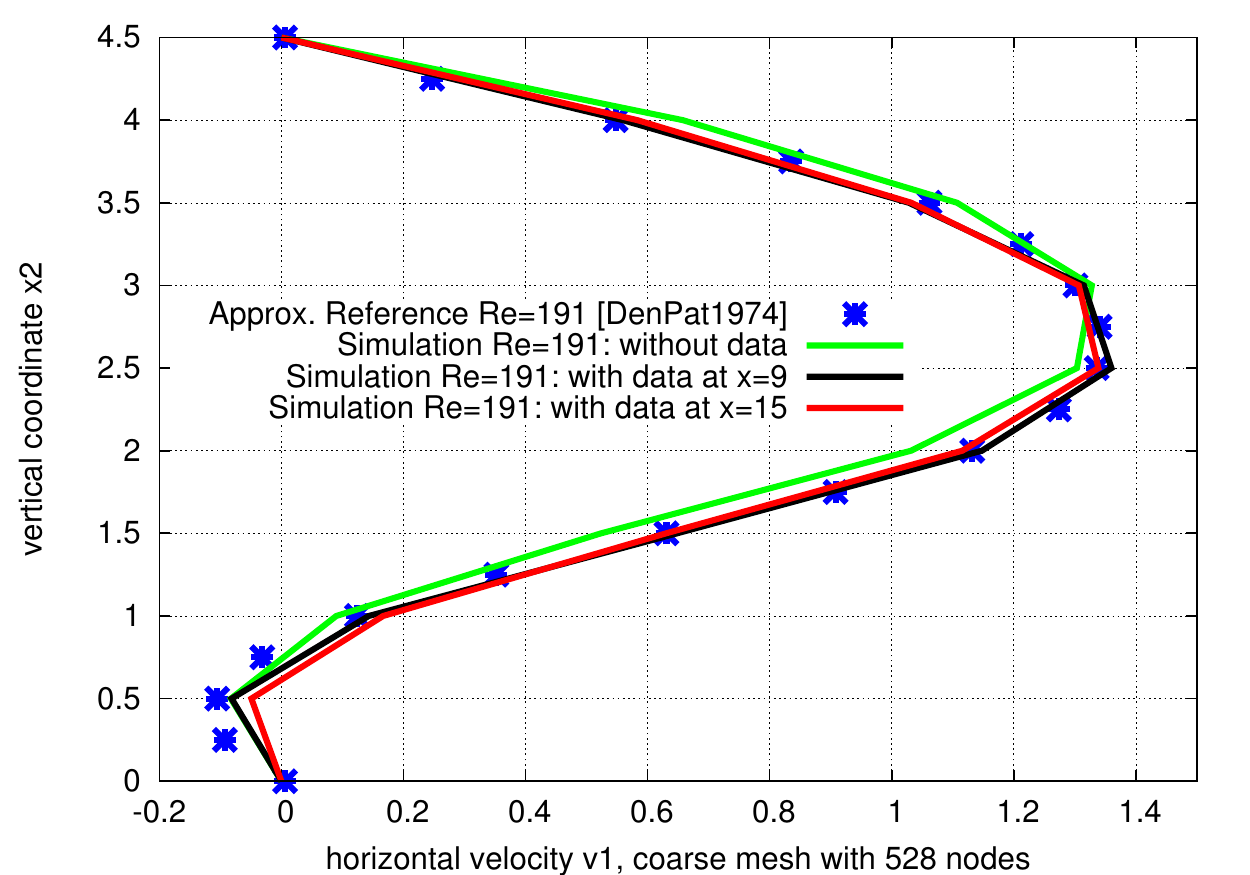}
}\\
\subfigure[$x=9$]{
\includegraphics[width=0.47\textwidth]{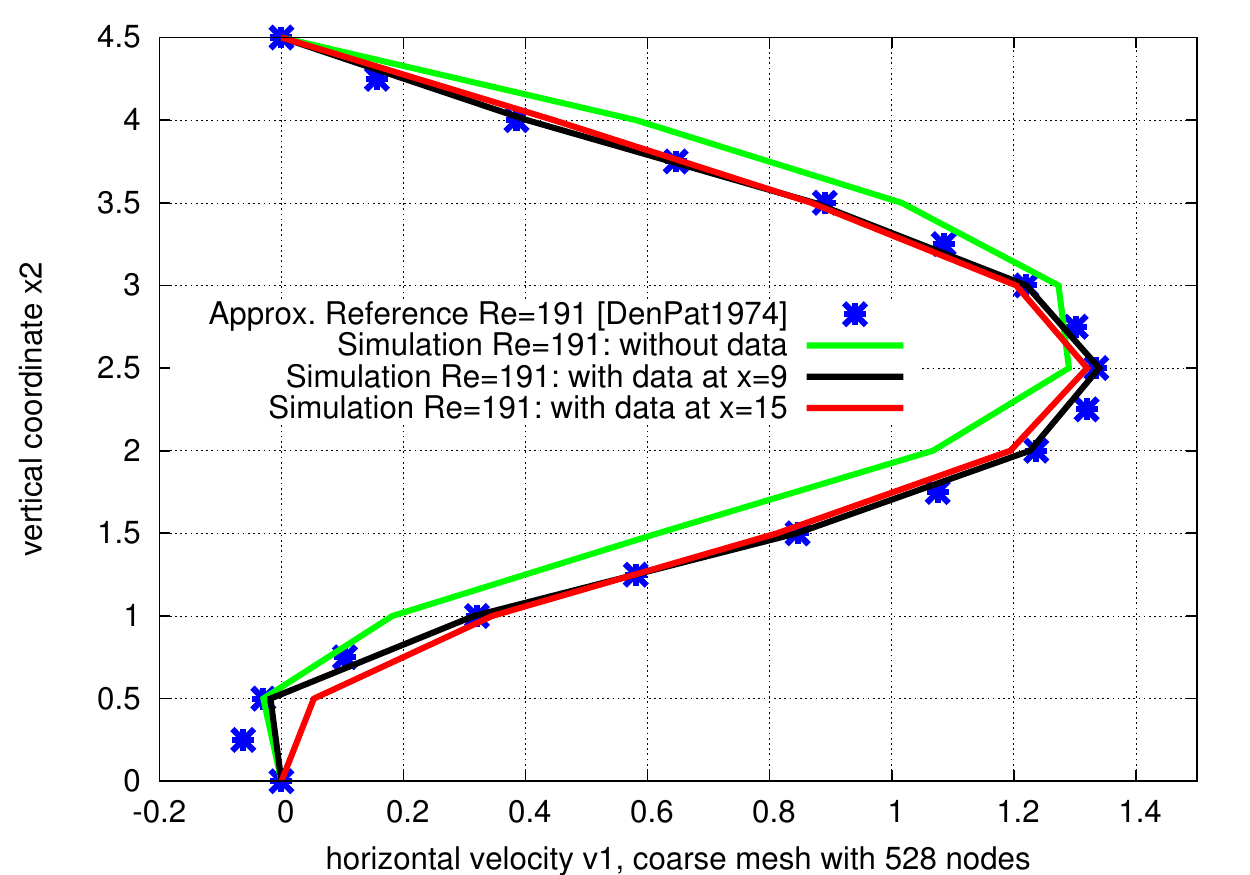}
}
\subfigure[$x=12$]{
\includegraphics[width=0.47\textwidth]{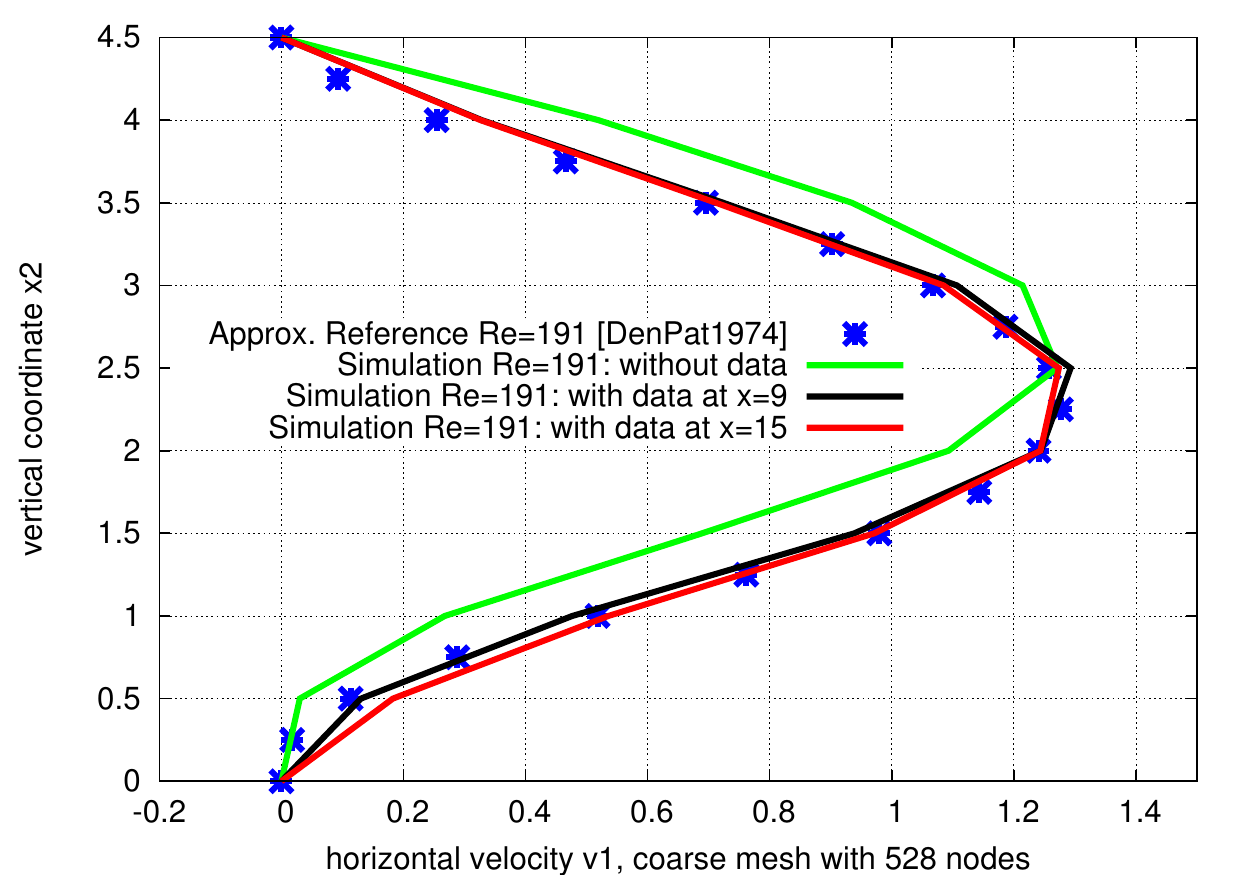}
}
\caption{Horizontal velocity profiles.  Coarse grid, experimental data at $x=9$ or $x=15$.}
\label{velodata1}
\end{figure}

The experimental data is included in the form of a velocity profile at either
$x_d = 9$ or $x_d = 15$, see Figure~\ref{velofielddata}.  The former reversed
flow near the wall; at the latter station the flow in the experiment is already
uni-directional.  Figure~\ref{velodata1} shows the results for the coarse mesh
without and with experimental data as well as the experimental reference
solution of ~\cite{DenPat1974} at $x=3$, $x=6$, $x=9$ and $x=12$.

It can be seen that at all locations a substantial reduction in the error is
achieved, not only at the location of the added data, but in the entire domain.
The improvement is most noticeable for $x_d=9,12$, where the original
discretization error is largest (due to mass loss and accumulation of errors
from upstream), and where the data has the most influence.  In
Figure~\ref{velodata1}(c) for $x=9$, and data at the same location $x_d=9$, we
can see the local influence of the data term.  The solution is prescribed mainly by
the data at this point, and follows it closely.  Note that due to the extremely
coarse grid cells, the recirculation region at this location is almost not
resolved in the data set.  This limits the improvement of the approximation of
this small feature.  At $x_d=15$ the data contains no recirculation region, yet an
improvement in the location of the maximum velocity at all $x$-stations is still
observed.

%
% COARSE

\begin{figure}[ht]
\centering
\subfigure[$x=3$]{
\includegraphics[width=0.47\textwidth]{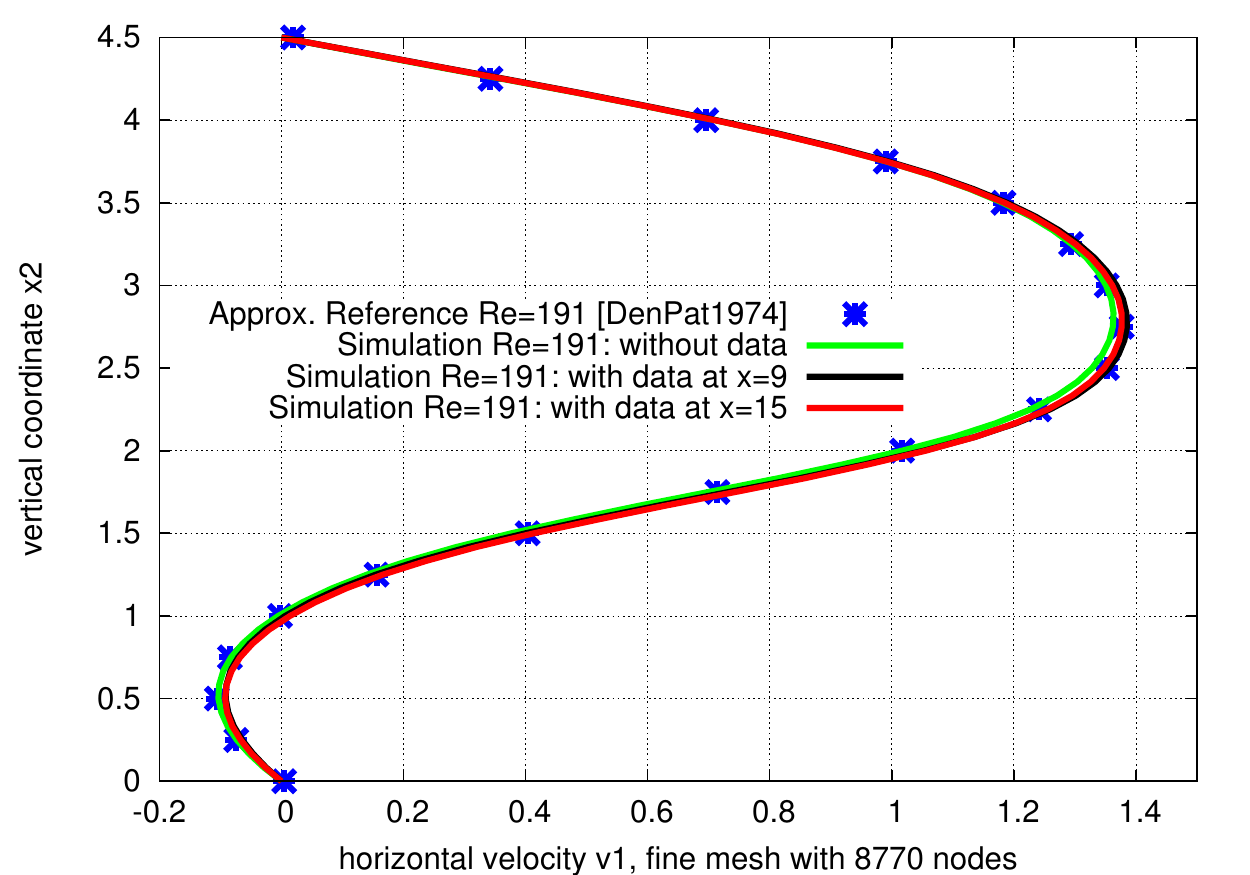}
}
\subfigure[$x=6$]{
\includegraphics[width=0.47\textwidth]{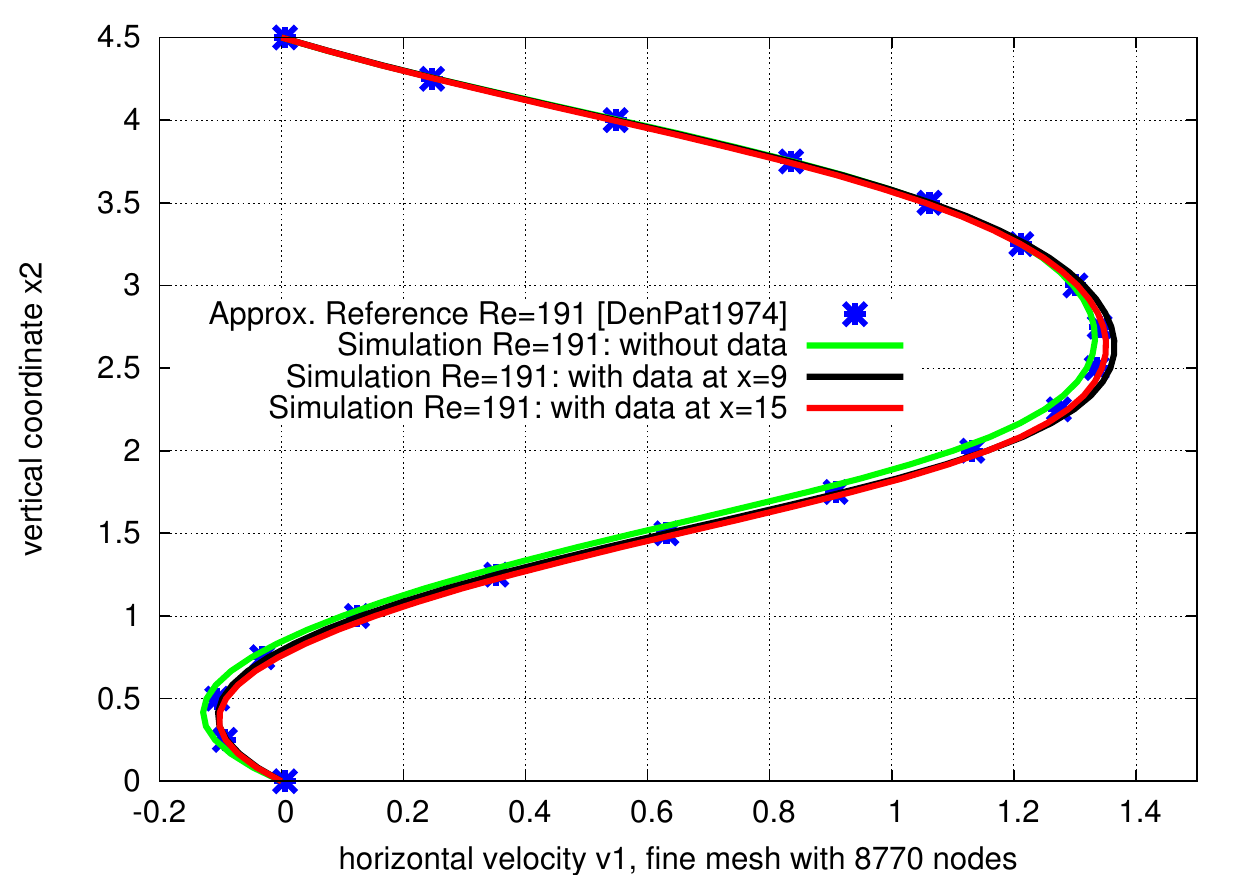}
}\\
\subfigure[$x=9$]{
\includegraphics[width=0.47\textwidth]{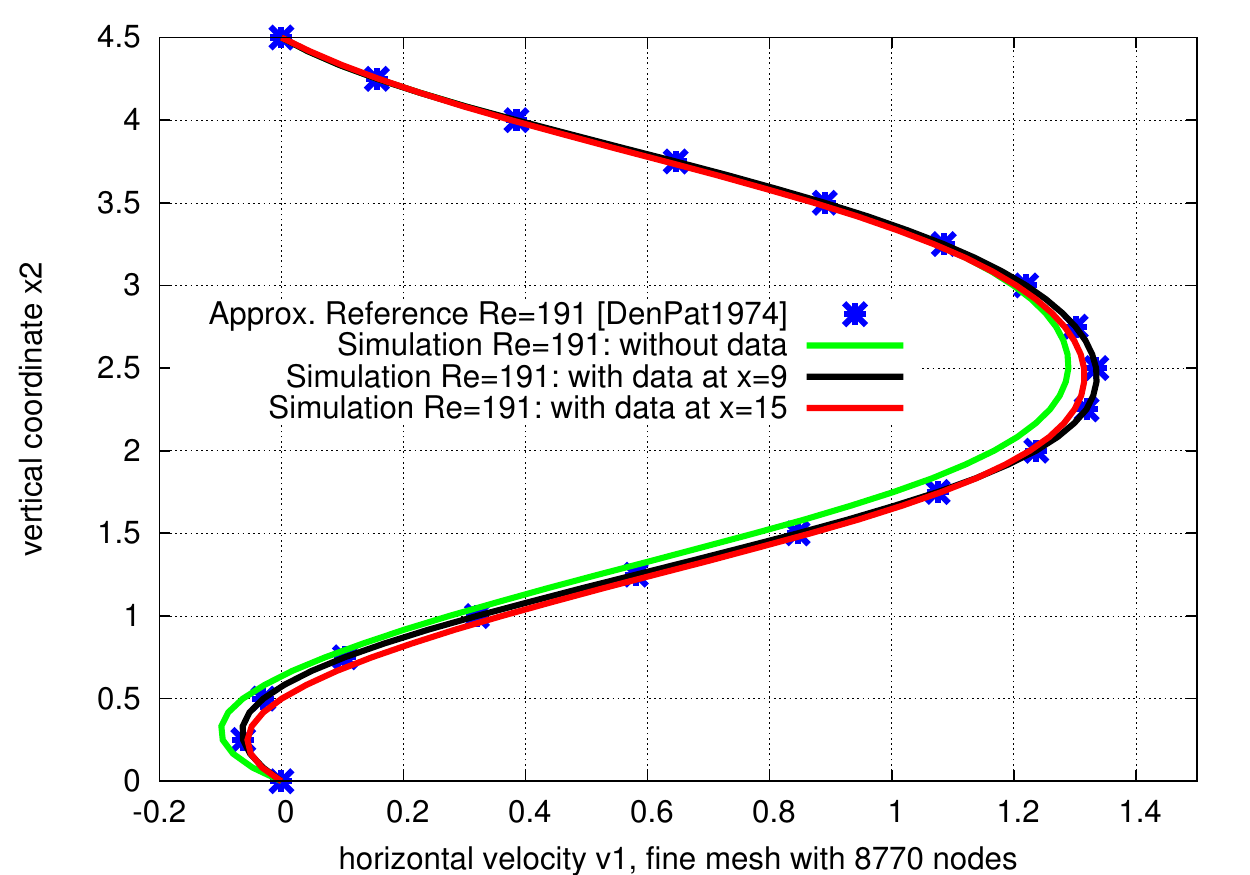}
}
\subfigure[$x=12$]{
\includegraphics[width=0.47\textwidth]{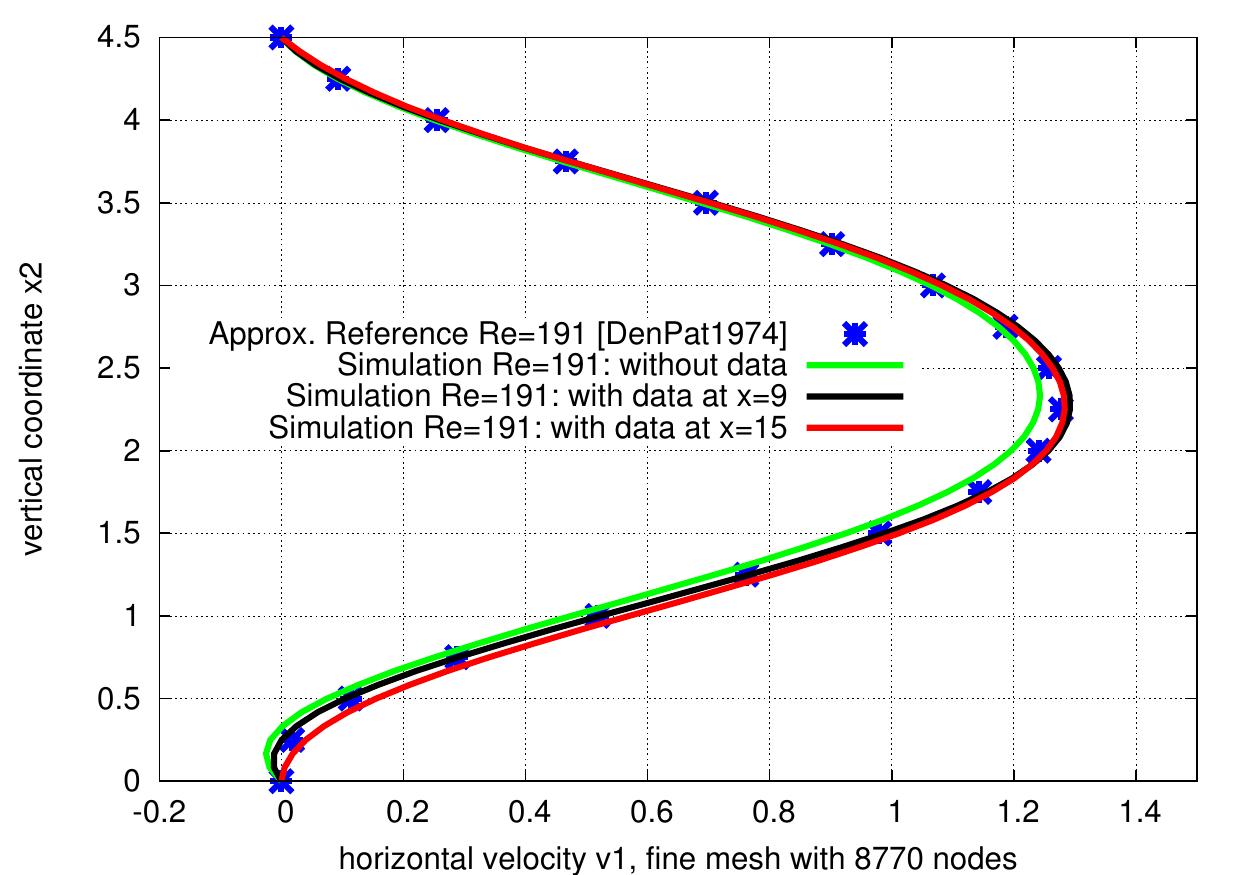}
}
\caption{Horizontal velocity profiles.  Fine grid, experimental data at $x=9$ or $x=15$.}
\label{velodata3}
\end{figure}

For the fine grid in Figure~\ref{veloexp1} there is also a defect, though in
this case it is not primarily due to discretization error.  Nonetheless we apply the
same procedure as for the coarse grid, with data introduced at the same
locations in the hope that this unidentified error will be reduced, see
Figure~\ref{velodata3}.  Since the fine grid has more nodes in the
$y$-direction, more data is used, in particular the reversed-flow region at
$x_d=9$ is resolved.  Again we see reduction in error across the full domain for
data introduced at both $x_d=9$ and $x_d=15$.  This time, not only the location and
strength of the peak is the benefactor, but also the reproduction of the
circulation region.
%
% FINE
%

%%%%%%%%%%%%%%%%%%%%%%%%%%%%%%%%%%%%%%%%%%%%%%%%%%%%%%%%%%%%%%%%%%%%%%%%%%%%%%%%
\subsection{Correcting Unknown and Uncertain BCs with Experimental Data}
\label{s_bcerror}
Reducing discretization error can be achieved most readily with mesh refinement.
However in Section~\ref{s_nodata} it was seen that for a reasonable mesh
resolution, error in the inflow BC dominated the total error balance.  Such
errors can not be reduced without reference to the experiment.  Consider the
situation in which the inflow conditions to the computational domain are
unmeasured.  This frequently occurs in experiments in fluids due to issues of
measurement device access, for example, optical access in PIV.

In order to imitate this situation we consider two possible states of knowledge
regarding the horizontal inflow boundary condition.  Firstly we assume
absolutely no information regarding the inflow velocity is available, secondly
we assume that the inflow is known to be roughly parabolic (as follows from
basic flow theory).  The experimental data is included at
$x=9$, see Section~\ref{s_example2}.  Furthermore, we assume zero vertical
velocity at the inflow in both cases.  The setup and material parameters of the
backward facing step at Reynolds number of $Re=191$ are the same as in the
previous section.

%%%%%%%%%%%%%%%%%%%%%%%%%%%%%%%%%%%%%%%%%%%%%%%%%%%%%%%%%%%%%%%%%%%%%%%%%%%%%%%%
\subsubsection{Simulation with fully unknown inflow conditions}
\label{s_bcerror1}
\begin{figure}[ht]
\centering
\subfigure[$x=-2$]{
\includegraphics[width=0.47\textwidth]{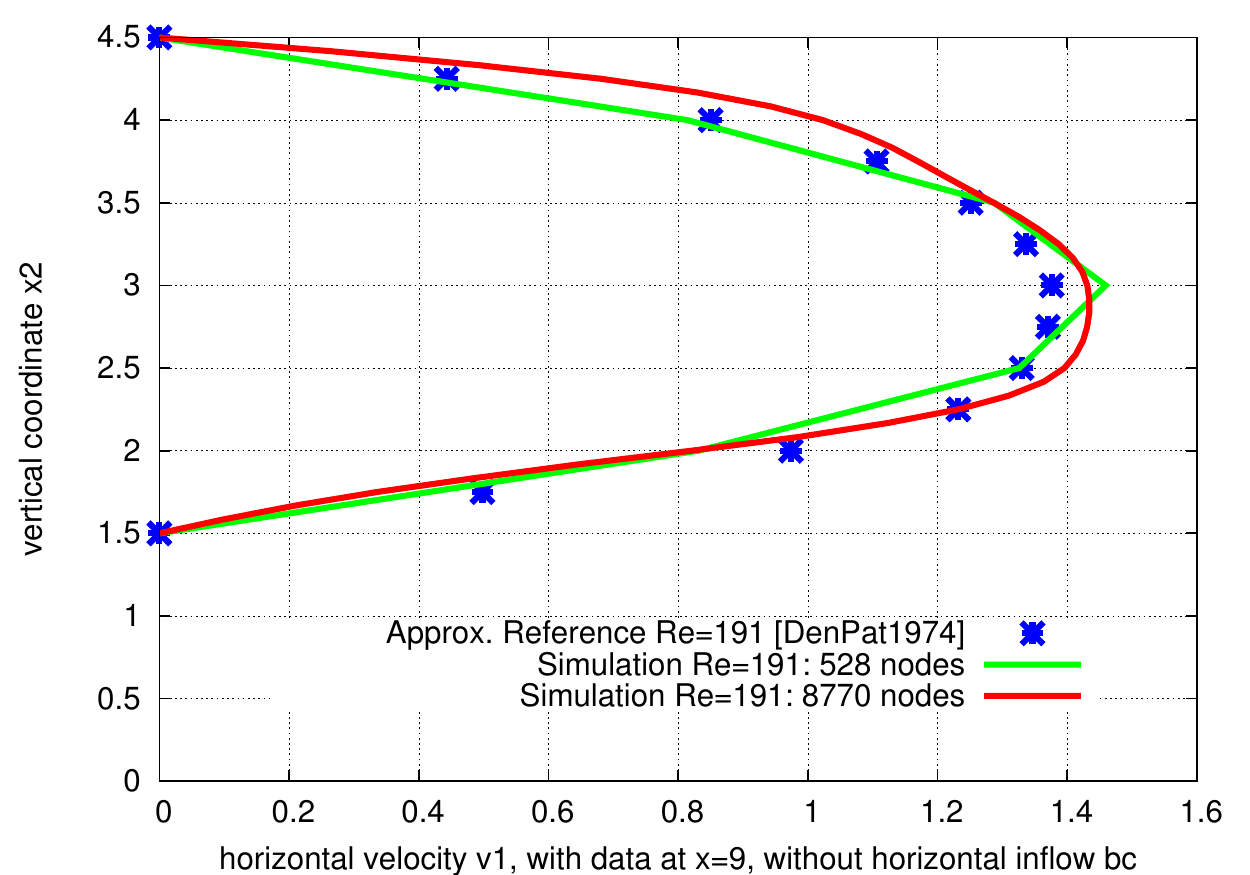}
}
\subfigure[$x=3$]{
\includegraphics[width=0.47\textwidth]{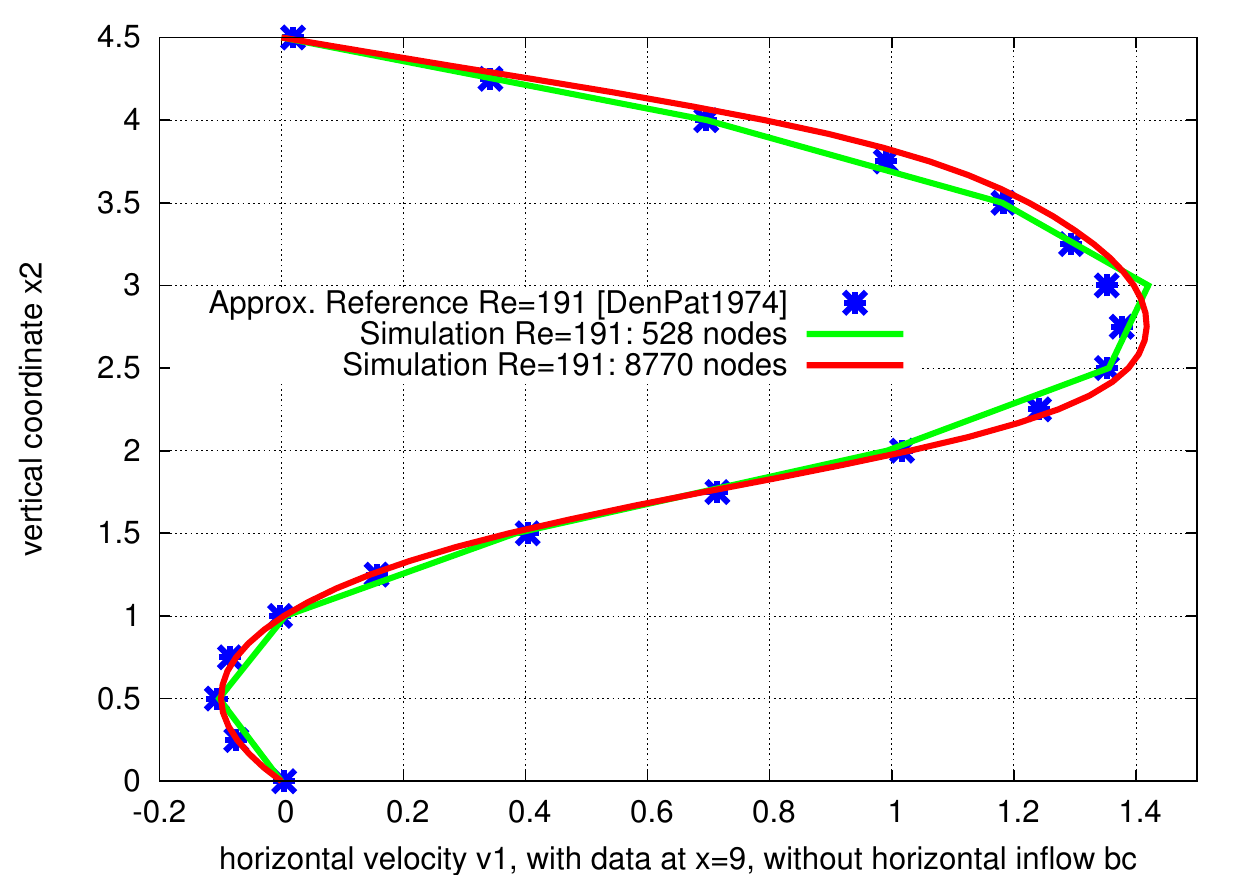}
}\\
\subfigure[$x=6$]{
\includegraphics[width=0.47\textwidth]{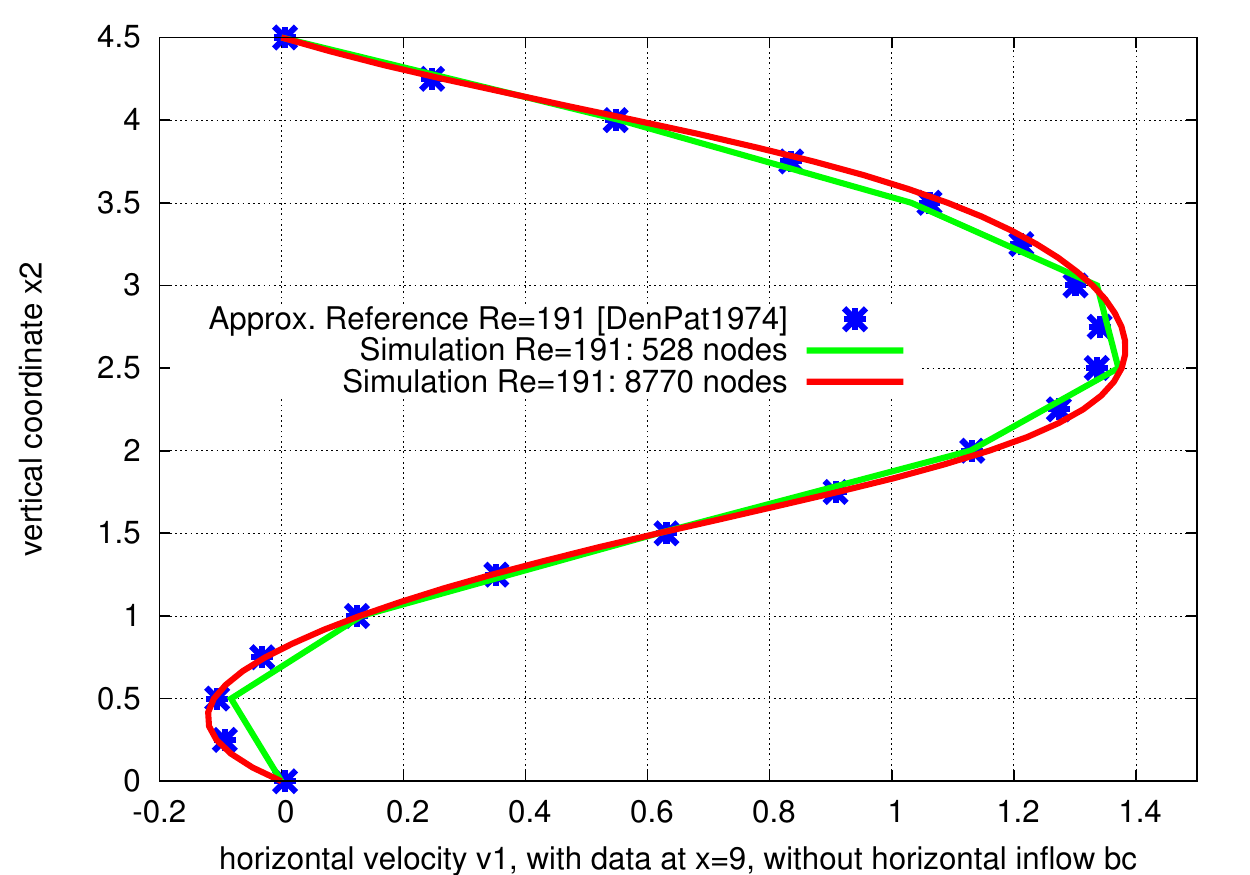}
}
\subfigure[$x=12$]{
\includegraphics[width=0.47\textwidth]{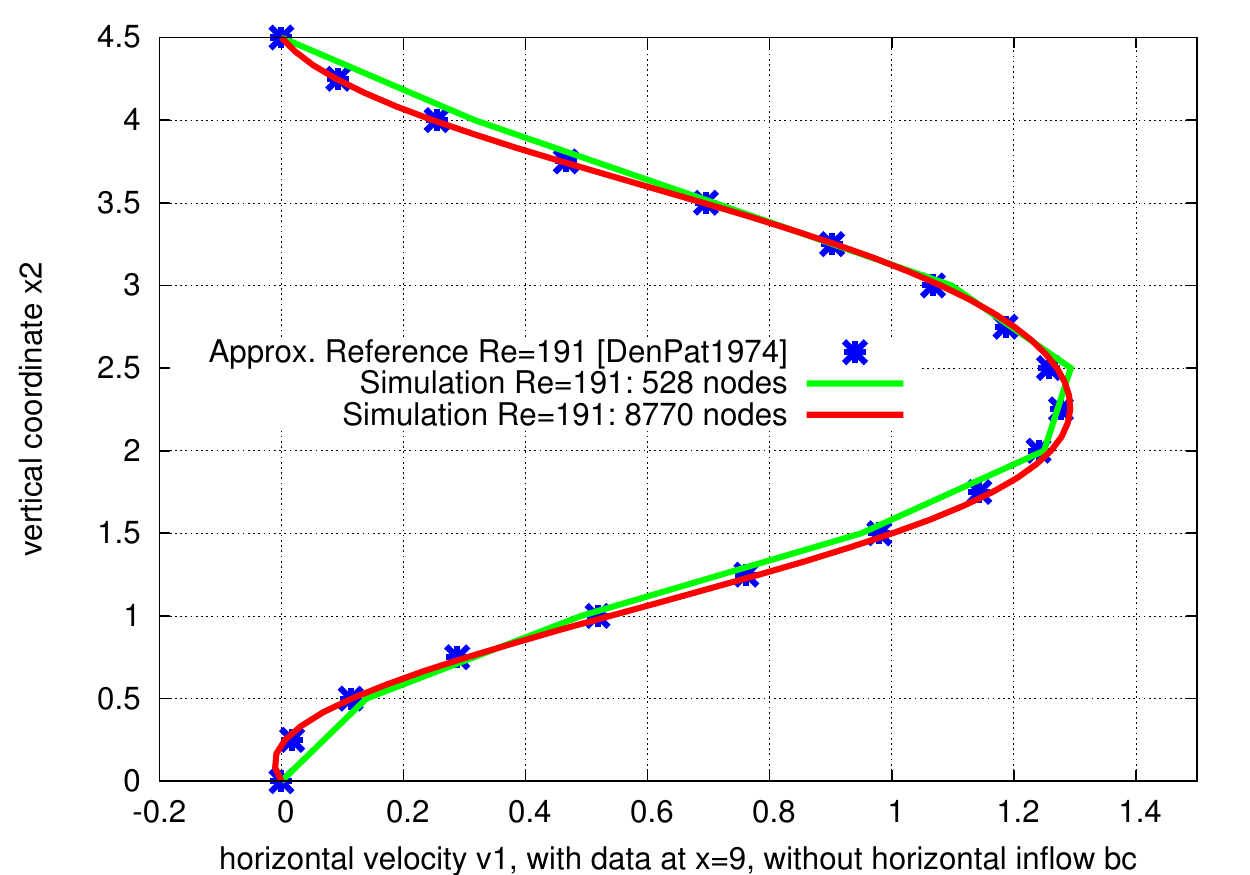}
}
\caption{Horizontal velocity profiles.  No inflow condition specified,
experimental data at $x_d=9$.}
\label{velonobc1}
\end{figure}

Here we consider the situation in which we assume {\it no} information about the inflow
condition.  We simply omit from the discretization all horizontal inflow BCs
(weak and strong).  This type of boundary value problem is a classic inverse
problem.  The results for the discretizations of 528 and 8770 nodes as well as
the reference solution from the experiment at the inflow at $x=3$, $x=6$ and
$x=12$ are depicted in Figure~\ref{velonobc1}.  We leave the results at $x=9$
out because they are identical to the reference solution as in
Section~\ref{s_example2}.

From Figure~\ref{velonobc1}(a) it is clear that the procedure
reproduces a physically reasonable inflow profile, that matches
roughly the reference profile from the experiment.  Most of the error
is related to an overestimation of the maximum velocity and the mass
flow.  Such an error could be due to the discretization losing some
mass in the channel.  When solving the inverse problem the method
therefore injects extra mass to account for this loss.  The closer we
get to $x=9$ the closer the results become to the experimental
reference solution.  It can be seen that our proposed approach is
capable of solving problems with unknown BCs in a quite satisfying
manner.

\subsubsection{Simulation with weak parabola inflow conditions}
\label{s_bcerror2}
\begin{figure}[ht]
\centering
\subfigure[$x=-2$]{
\includegraphics[width=0.47\textwidth]{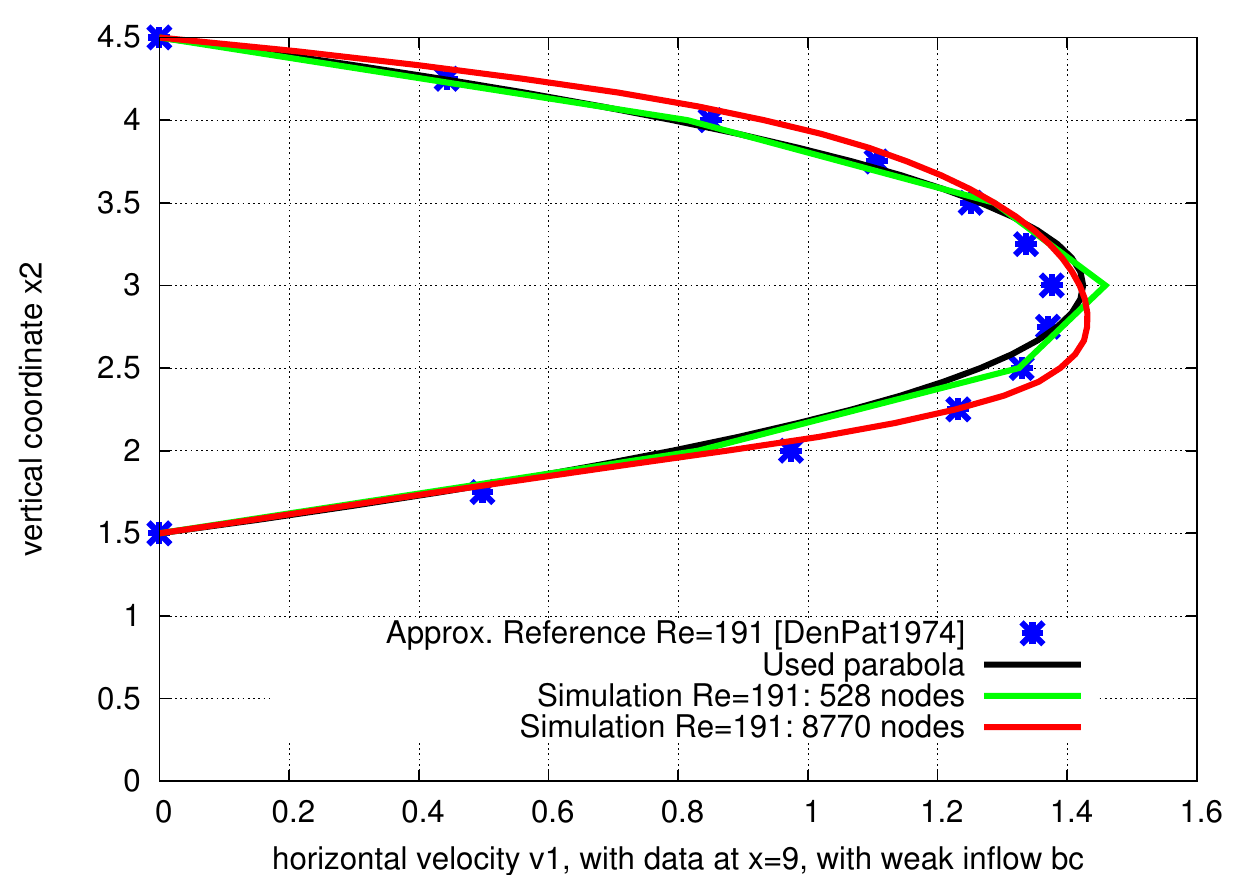}
}
\subfigure[$x=3$]{
\includegraphics[width=0.47\textwidth]{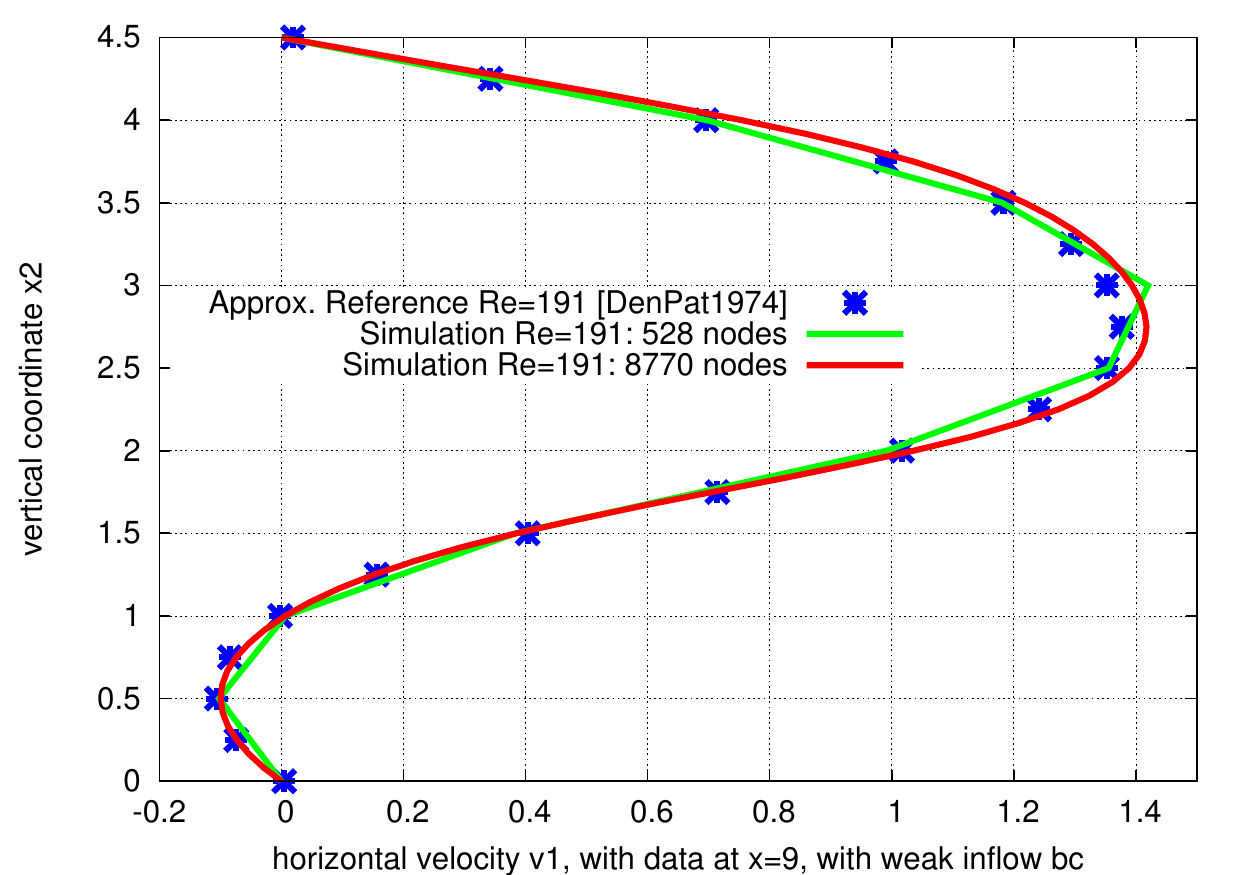}
}\\
\subfigure[$x=6$]{
\includegraphics[width=0.47\textwidth]{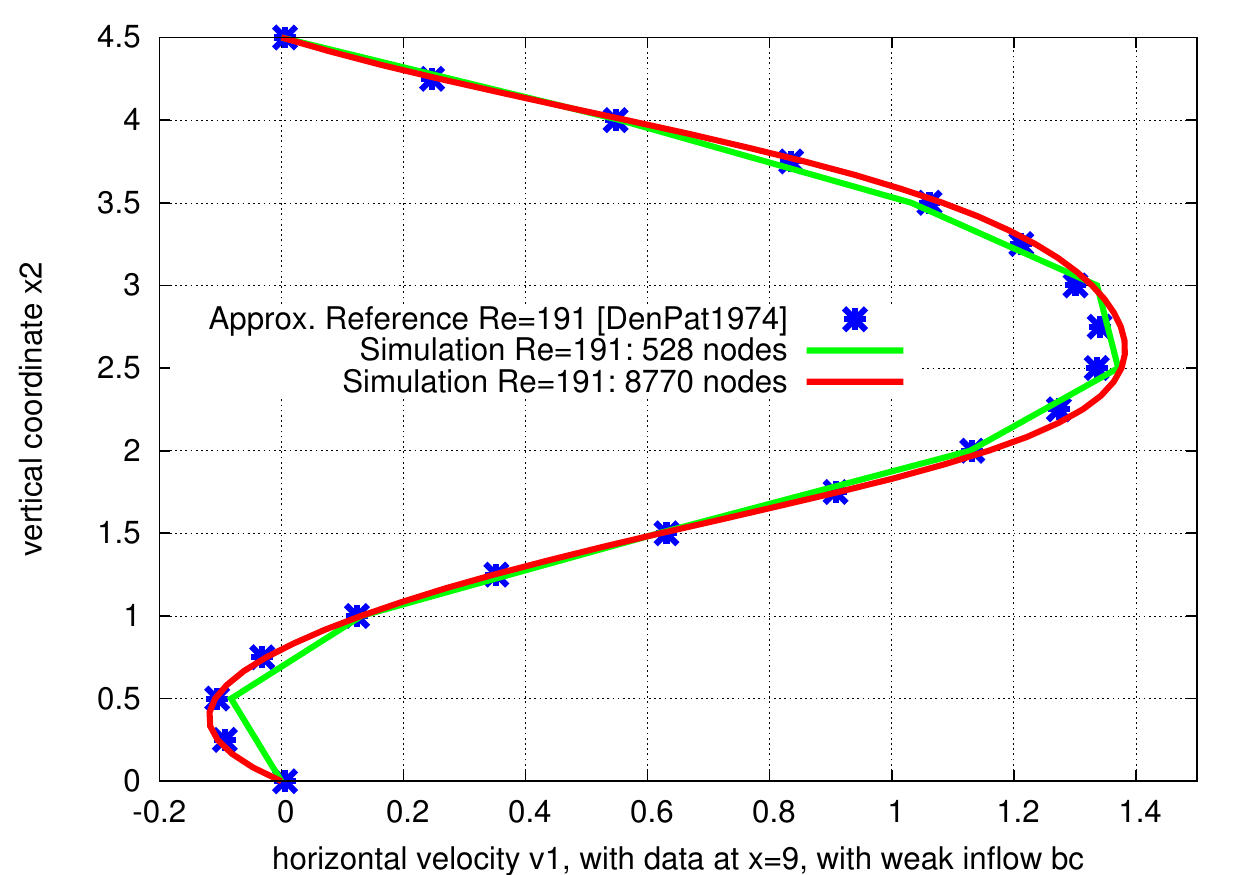}
}
\subfigure[$x=12$]{
\includegraphics[width=0.47\textwidth]{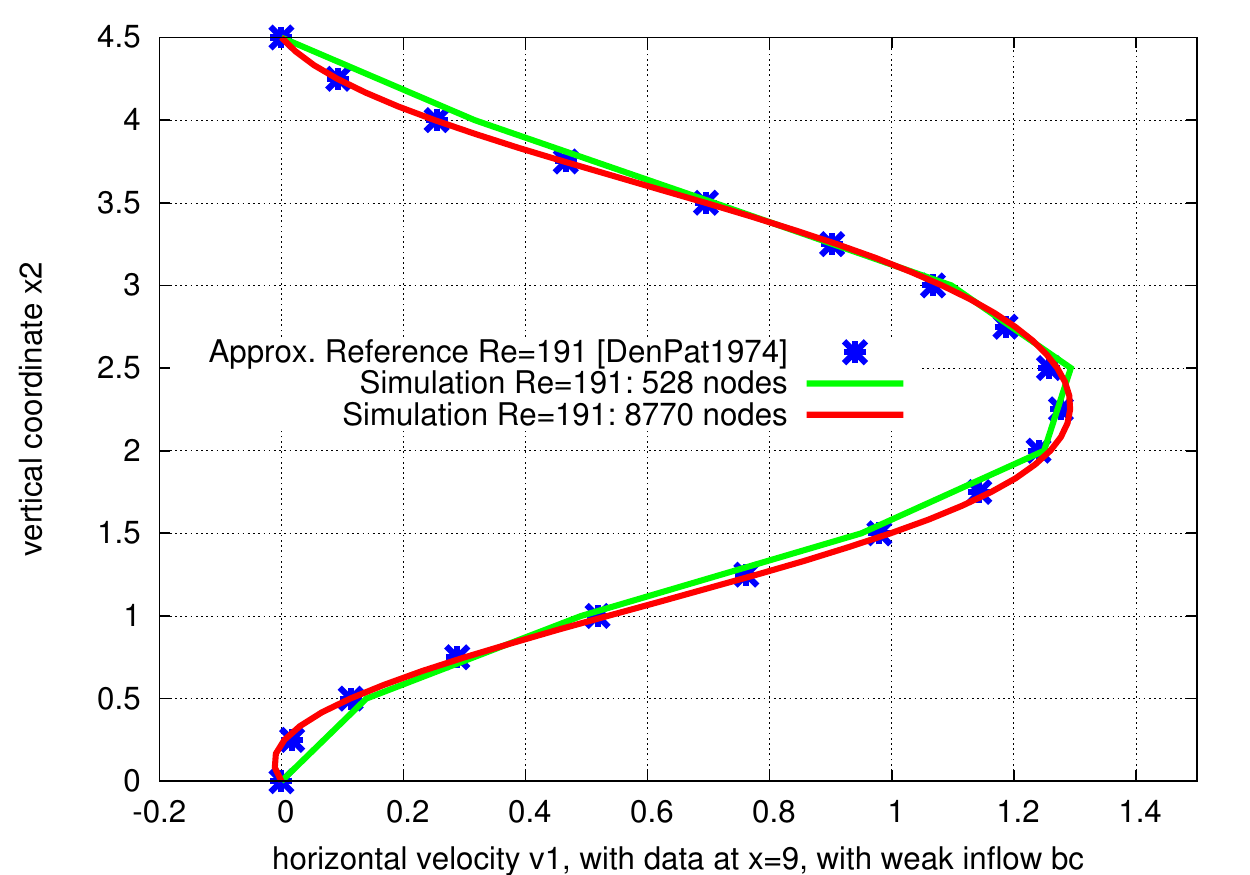}
}
\caption{Horizontal velocity profiles.  Weakly enforced parabolic inflow, experimental data at $x_d=9$.}
\label{velo451}
\end{figure}

Now we try to improve on the results of the previous section by
incorporating some inexact knowledge of the horizontal inflow velocity
-- namely that it is roughly parabolic with an estimated maximum
velocity.  We employ the parabola from Section~\ref{s_nodata} but in
order to accommodate the uncertainty in this assumption we include the
parabolic BC in weak form with a coefficient of $\omega=0.01$.  
The implementation of weak boundary conditions is realized in similar
form to the assimilation data term.  As in the
previous section we omit the results at $x=9$ and consider the results
at the inflow ($x=-2$), and at $x=3, 6, 12$ for the fine and coarse
discretizations.  In Figure~\ref{velo451} the results for the
horizontal velocity profiles with weak parabolic inflow boundary
conditions are depicted.

Looking first at Figure~\ref{velo451}(a), we see an improvement in the
quality of the velocity profile on the fine grid at the inflow.  Most
noticeably, the bump present at $y=4$ in Figure~\ref{velonobc1}(a) is
removed.  However also on the lower side of the profile, the
experimental match is slightly better, and the overall mass flow is
substantially corrected.  Downstream the effect is less pronounced, and
very little difference with respect to the no-BC case is visible.
Overall we see that using approximate information at the inflow in the
form of a weak BC, gives an improved solution compared to using no
information there.

Note that these results are contingent on a reasonable choice of the
weak BC weighting.  Choosing the BC weighting too small will
result the weak BC having negligible influence on the solution (giving
the results of Section~\ref{s_bcerror1}), too large and the inflow
condition will be forced to match the (incorrect) black parabola in
Figure~\ref{velo451}(a).

%%%%%%%%%%%%%%%%%%%%%%%%%%%%%%%%%%%%%%%%%%%%%%%%%%%%%%%%%%%
\section{Conclusions}
\label{s_conclusions}
It has been shown how a modified LSFEM solver, combined with
experimental data can:
\begin{itemize}
\item Correct the effects of discretization error,
\item Correct inaccurate boundary conditions,
\item Approximate fully unknown inflow conditions,
\end{itemize}
%
% Data saves the day!
at a cost no greater than that of a standard LSFEM solve.  In numerical experiments
on a backward-facing step the introduction of experimental velocity
profiles at one location has substantially improved the match with the
experiment both up- and downstream.  This has applications in many
situations in which direct measurement of quantities of interest is
impractical, but in which measurement of related quantities are easy.

The number one obstacle to the adoption of this technique in practice, for
example to augment gappy PIV data in the style of Sciacchitano et
al.~\cite{SchDwiSca2012}, is the limited capabilities of current LSFEM codes.
To be useful in postprocessing experiments in incompressible aerodynamics the
method must be able to treat high Reynolds numbers, use RANS turbulence
modelling, with good boundary-layer resolution, low dissipation and little
mass-loss.  It is hoped that with further numerical developments such as the
introduction of higher-order elements, improved first-order system choices, 
and local mesh refinement~\cite{Becker1998,Dwi2008} -- applications of LSFEM to fluids will
become more common.  The approach described in this paper can nonetheless be
employed for other physical problems where efficient LSFEM solvers already
exist, e.g.\ in magnetohydrodynamics~\cite{AdlManMcc2010}.  Furthermore the
ability to assimilate experimental measurements at no additional cost, may make
LSFEM competitive in applications in which it would not be competitive as a
stand-alone solver.

%%%%%%%%%%%%%%%%%%%%%%%%%%%%%%%%%%%%%%%%%%%%%%%%%%%%%%%%%%%
\bibliographystyle{elsarticle-num}
\bibliography{refs}

\begin{thebibliography}{10}
\expandafter\ifx\csname url\endcsname\relax
  \def\url#1{\texttt{#1}}\fi
\expandafter\ifx\csname urlprefix\endcsname\relax\def\urlprefix{URL }\fi
\expandafter\ifx\csname href\endcsname\relax
  \def\href#1#2{#2} \def\path#1{#1}\fi

\bibitem{Rag2012}
D.~Ragni, {PIV}-based loads determination in aircraft propellers, Ph.D. thesis,
  TU Delft (2012).

\bibitem{ScaMoo2012}
F.~Scarano, P.~Moore, An advection-based model to increase the temporal
  resolution of {PIV} time series, Experiments in Fluids 52 (2012) 919--933.

\bibitem{SchDwiSca2012}
A.~Sciacchitano, R.~Dwight, F.~Scarano, Navier-{S}tokes simulations in gappy
  {PIV} data, Experiments in Fluids.

\bibitem{ConAni2012}
E.~Constantinescu, M.~Anitescu, Physics-based covariance models for {G}aussian
  processes with multiple outputs, International Journal on Uncertainty
  Quantification.

\bibitem{RagVouSca2011}
D.~Ragni, B.~van Oudheusden, F.~Scarano, Non-intrusive aerodynamic loads
  analysis of an aircraft propeller blade, Experiments in Fluids 51 (2011)
  361--371.

\bibitem{Jia1998}
B.-N. Jiang, The Least-Squares Finite Element Method: {T}heory and Applications
  in Computational Fluid Dynamics and Electromagnetics, 1st Edition, Springer,
  Berlin, 1998.

\bibitem{BocGun2009}
P.~B. Bochev, M.~D. Gunzburger, Least-Squares Finite Element Methods, 1st
  Edition, Vol. 166 of Applied Mathematical Sciences, Springer, New York, 2009.

\bibitem{HeyManMcc2010}
J.~Heys, T.~Manteuffel, S.~McCormick, M.~Milano, J.~Westerdale, M.~Belohlavek,
  Weighted least-squares finite elements based on particle imaging velocimetry
  data, Journal of Computational Physics 229~(1) (2010) 107--118.
\newblock \href {http://dx.doi.org/doi: 10.1016/j.jcp.2009.09.016}
  {\path{doi:doi: 10.1016/j.jcp.2009.09.016}}.

\bibitem{Dwi2010}
R.~Dwight, Bayesian inference for data assimilation using least-squares finite
  element methods, IOP Series: Materials Science and Engineering 10~(1) (2010)
  222--232.
\newblock \href {http://dx.doi.org/doi:10.1088/1757-899X/10/1/012224}
  {\path{doi:doi:10.1088/1757-899X/10/1/012224}}.

\bibitem{CaiLeeWan:2004:lsm}
Z.~Cai, B.~Lee, P.~Wang, Least-squares methods for incompressible newtonian
  fluid flow: linear stationary problems, SIAM Journal on Numerical Analysis 42
  (2004) 843--859.

\bibitem{Eve2003}
G.~Evensen, The ensemble {K}alman filter: {T}heoretical formulation and
  practical implementation, Ocean Dynamics 53 (2003) 343--367.

\bibitem{EveLee2000}
G.~Evensen, P.~van Leeuwen, An ensemble {K}alman smoother for nonlinear
  dynamics, Monthly Weather Review 128 (2000) 1852--1867.

\bibitem{LegAlo2004}
P.~LeGresley, J.~Alonso, Improving the Performance of Design Decomposition
  Methods with {POD}, American Institute of Aeronautics and Astronautics, 2004.

\bibitem{ZimGoe2012}
R.~Zimmermann, S.~Görtz, Non-linear reduced order models for steady turbulent
  aerodynamics, Royal Aerospace Society Journal.

\bibitem{HayImaFun2010}
T.~Hayase, K.~Imagawa, K.~Funamoto, , A.~Shirai, Stabilization of
  measurement-integrated simulation by elucidation of destabilizing mechanism,
  Journal of Fluid Science and Technology 5~(3) (2010) 632--647.

\bibitem{ImaHay2010}
K.~Imagawa, T.~Hayase, Eigenvalue analysis of linearized error dynamics of
  measurement integrated flow simulation, Computers and Fluids 39~(10) (2010)
  1796--1803.

\bibitem{Jia:1998:tls}
B.-N. Jiang, The Least-Squares Finite Element Method, Springer-Verlag, Berlin,
  1998.

\bibitem{BocGun:2009:lsf}
P.~Bochev, M.~Gunzburger, Least-Squares Finite Element Methods, 1st Edition,
  Springer-Verlag, New York, 2009.

\bibitem{KayMat:2005:lsf}
O.~Kayser-Herold, H.~Matthies, Least-squares {FEM}, literature review,
  Informatik-Bericht 2005-05, TU Braunschweig, Institut f{\"u}r
  Wissenschaftliches Rechnen.

\bibitem{pon:2003:lsv}
J.~P. Pontaza, Least-squares variational principles and the finite element
  method: theory, form, and model for solid and fluid mechanics, Ph.D. thesis,
  {T}exas {A\&M} {U}niversity (2003).

\bibitem{JiaCha:1990:lsf}
B.-N. Jiang, C.~Chang, Least-squares finite elements for the {S}tokes problem,
  Computer Methods in Applied Mechanics and Engineering 78 (1990) 297--311.

\bibitem{Boc:1997:aol}
P.~Bochev, Analysis of least-squares finite element methods for the
  {N}avier-{S}tokes equations, SIAM Journal on Numerical Analysis 34 (1997)
  1817--1844.

\bibitem{BelSur:1994:pls}
B.~Bell, K.~Surana, p-version least squares finite element formulation for
  two-dimensional, incompressible, non-newtonian isothermal and non-isothermal
  fluid flow, International Journal for Numerical Methods in Fluids 18 (1994)
  127--162.

\bibitem{BocGun:1995:lsm}
P.~Bochev, M.~Gunzburger, Least-squares methods for the
  velocity-pressure-stress formulation of the {S}tokes equations, Computer
  Methods in Applied Mechanics and Engineering 126 (1995) 267--287.

\bibitem{DinTsa:2003:ofo}
X.~Ding, T.~Tsang, On first-order formulations of the least-squares finite
  element method for incompressible flows, International Journal of
  Computational Fluid Dynamics 17 (2003) 183--197.

\bibitem{CaiManMcC:1997:fos}
Z.~Cai, T.~Manteuffel, S.~McCormick, First-order system least squares for the
  {S}tokes equation, with application to linear elasticity, SIAM Journal on
  Numerical Analysis 34 (1997) 1727--1741.

\bibitem{BocCaiManMcC:1998:aov}
P.~Bochev, Z.~Cai, T.~Manteuffel, S.~McCormick, Analysis of velocity-flux
  least-squares methods for the {N}avier-{S}tokes equations, part {I}, SIAM
  Journal on Numerical Analysis 35 (1998) 990--1009.

\bibitem{BocManMcC:1999:aov}
P.~Bochev, T.~Manteuffel, S.~McCormick, Analysis of velocity-flux least squares
  methods for the {N}avier-{S}tokes equations, part {II}, SIAM Journal on
  Numerical Analysis 36 (1999) 1125--1144.

\bibitem{DeaGun:1998:irt}
J.~M. Deang, M.~D. Gunzburger, Issues related to least-squares finite element
  methods for the {S}tokes equations, SIAM Journal on Scientific Computing
  20~(3) (1998) 878--906.

\bibitem{BolTha:2005:omc}
P.~Bolton, R.~W. Thatcher, On mass conservation in least-squares methods,
  Journal of Computational Physics 203 (2005) 287--304.

\bibitem{ChaNel:1997:lsf}
C.~L. Chang, J.~J. Nelson, Least-squares finite element method for the {S}tokes
  problem with zero residual of mass conservation, SIAM Journal on Scientific
  Computing 34~(2) (1997) 480--489.

\bibitem{Pon:2006:als}
J.~P. Pontaza, A least-squares finite element form for unsteady incompressible
  flows with improved velocity-pressure coupling, Journal of Computational
  Physics 217 (2006) 563--588.

\bibitem{ponred:2003:sls}
J.~P. Pontaza, J.~N. Reddy, Spectral/$hp$ least-squares finite element
  formulation for the {N}avier-{S}tokes equation, Journal of Computational
  Physics 190 (2003) 523--549.

\bibitem{ponred:2004:stc}
J.~P. Pontaza, J.~N. Reddy, Space-time coupled spectral/$hp$ least-squares
  finite element formulation for the incompressible {N}avier-{S}tokes equation,
  Journal of Computational Physics 197 (2004) 418--459.

\bibitem{ProGer:2006:mam}
M.~M.~J. Proot, M.~I. Gerritsma, Mass- and momentum conservation of the
  least-squares spectral element method for the {S}tokes problem, Journal of
  Scientific Computing 27 (2006) 389--401.

\bibitem{HeyLeeManMcC:2006:omc}
J.~J. Heys, E.~Lee, T.~A. Manteuffel, S.~F. McCormick, On mass-conserving
  least-squares methods, SIAM Journal on Scientific Computing 28~(3) (2006)
  1675--1693.

\bibitem{HeyLeeManMcC:2007:aal}
J.~Heys, E.~Lee, T.~Manteuffel, S.~McCormick, An alternative least-squares
  formulation of the navier-stokes equations with improved mass conservation,
  Journal of Computational Physics 226 (2007) 994--1006.

\bibitem{HeyLeeManMcCRug:2009:emc}
J.~Heys, E.~Lee, T.~Manteuffel, J.~McCormick, S.and~Ruge, Enhanced mass
  conservation in least-squares methods for navier-stokes equations, SIAM J.
  Sci. Comp. 31 (2009) 2303--2321.

\bibitem{Kay:2006:lsm}
O.~Kayser-Herold, Least-squares methods for the solution of fluid-structure
  interaction problems, Dissertation, Technische Universit{\"a}t Braunschweig,
  Carl-Friedrich-Gauss-Fakult{\"a}t f{\"u}r Mathematik und Informatik (2006).

\bibitem{SchNicSerNisOuaSchTur:2016:acs}
A.~Schwarz, M.~Nickaeen, S.~Serdas, C.~Nisters, A.~Ouazzi, J.~Schr{\"{o}}der,
  S.~Turek, {A Comparative Study of Mixed Least Squares FEMs for the
  Incompressible Navier-Stokes Equations}, International journal of
  Computational Science and Engineering in print.

\bibitem{RavTho:1977:amf}
P.~A. Raviart, J.~M. Thomas, A mixed finite element method for 2-nd order
  elliptic problems. {M}athematical aspects of finite element methods, Lecture
  Notes in Mathematics, Springer-Verlag New York (1977) 292--315.

\bibitem{BreFor:1991:mah}
F.~Brezzi, M.~Fortin, Mixed and Hybrid Finite Element Methods, Springer-Verlag,
  NewYork, 1991.

\bibitem{KenOha2001}
M.~Kennedy, A.~O'Hagan, Bayesian calibration of computer models (with
  discussion), Journal of the Royal Statistical Society, Series B. 63 (2001)
  425--464.

\bibitem{NguPer2016}
N.~C. Nguyen, H.~Men, R.~M. Freund, J.~Peraire, Functional regression for state
  prediction using linear pde models and observations, SIAM Journal on
  Scientific Computing 38~(2) (2016) B247--B271.
\newblock \href {http://dx.doi.org/10.1137/14100275X}
  {\path{doi:10.1137/14100275X}}.

\bibitem{DenPat1974}
M.~Denham, M.~Patrick, Laminar flow over a downstream-facing step in a
  two-dimensional flow channel, Transactions of the Institution of Chemical
  Engineers 52 (1974) 361--367.

\bibitem{BarFon2002}
R.~Barber, A.~Fonty, A numerical study of laminar flow over a confined
  backward-facing step using a novel viscous-splitting vortex algorithm, in:
  Proc. 4th GRACM Congress on Computational Mechanics, Patra, 2002, pp. 1--8.

\bibitem{DebDwiBij2011b}
J.~de~Baar, R.~Dwight, H.~Bijl, Fast maximum likelihood estimate of the kriging
  correlation range in the frequency domain, in: International Association for
  Mathematical Geosciences (IAMG) Conference, Salzburg, 2011.

\bibitem{Oha1992}
A.~O'Hagan, Some {B}ayesian numerical analysis, Bayesian Statistics 4 (1992)
  345--363.

\bibitem{WikBer2007}
C.~K. Wikle, L.~M. Berliner, A {B}ayesian tutorial for data assimilation,
  Physica D: Nonlinear Phenomena 230~(1-2) (2007) 1--16.

\bibitem{Becker1998}
R.~Becker, R.~Rannacher, Weighted a posteriori error control in {FE} methods,
  in: Proceedings of ENUMATH-97, Heidelberg, World Scientific Publishing,
  Singapore, 1998.

\bibitem{Dwi2008}
R.~Dwight, Heuristic a posteriori estimation of error due to dissipation in
  finite volume schemes and application to mesh adaptation, Journal of
  Computational Physics 227~(5) (2008) 2845--2863.
\newblock \href {http://dx.doi.org/doi:10.1016/j.jcp.2007.11.020}
  {\path{doi:doi:10.1016/j.jcp.2007.11.020}}.

\bibitem{AdlManMcc2010}
J.~Adler, T.~Manteuffel, S.~McCormick, J.~Ruge, First-order system least
  squares for incompressible resistive magnetohydrodynamics, SIAM J. Sci.
  Comput. (SISC) 32~(1) (2010) 229--248.

\end{thebibliography}
\end{document}